\documentclass{aa}  
\usepackage{color}

\usepackage{graphicx}
\usepackage{txfonts,longtable}
\newcounter{subfig}
\def\bz{$\langle$B$_z\rangle$}
\def\nz{$\langle$N$_z\rangle$}

\begin{document}

   \title{B fields in OB stars (BOB): Detection of a magnetic field in
   the\\ He-strong star CPD\,$-$57$^{\circ}$\,3509\thanks{Based on observations made with
   ESO Telescopes at the La Silla Paranal Observatory under programme 
   ID 191.D-0255(C,E,F,G) and 171.D-0237(A).}\fnmsep
   \thanks{Figures~\ref{fig:fitsMET}-\ref{fig:fitsMET2} and Table~\ref{tab:linelist} are only available in electronic form via
   {\tt http://www.edpsciences.org}.}}

   \subtitle{}

   \author{N. Przybilla\inst{\ref{inst1}} 
   \and L. Fossati\inst{\ref{inst2},\ref{inst3}} 
   \and S. Hubrig\inst{\ref{inst4}}
   \and M.-F. Nieva\inst{\ref{inst1}}
   \and S.P. J\"arvinen\inst{\ref{inst4}}
   \and N. Castro\inst{\ref{inst3}} 
   \and M. Sch\"oller\inst{\ref{inst5}} 
   \and I. Ilyin\inst{\ref{inst4}}
   \and K. Butler\inst{\ref{inst6}}
   \and\\F.R.N. Schneider\inst{\ref{inst7}} 
   \and L.M. Oskinova\inst{\ref{inst8}}
   \and T. Morel\inst{\ref{inst9}} 
   \and N. Langer\inst{\ref{inst3}}
   \and A. de Koter\inst{\ref{inst10},\ref{inst11}}
   \and the BOB collaboration}

   \institute{Institut f\"ur Astro- und Teilchenphysik, Universit\"at
   Innsbruck, Technikerstr. 25/8, 6020 Innsbruck,
   Austria\\ \email{norbert.przybilla@uibk.ac.at}\label{inst1}
   \and
   Space Research Institute, Austrian Academy of Sciences,
   Schmiedlstr. 6, 8042 Graz, Austria\label{inst2}
   \and
   Argelander-Institut f\"ur Astronomie der Universit\"at Bonn,
   Auf dem H\"ugel 71, 53121 Bonn, Germany\label{inst3}
   \and
   Leibniz-Institut f\"ur Astrophysik Potsdam (AIP), An der 
   Sternwarte 16, 14482 Potsdam, Germany\label{inst4}
   \and
   European Southern Observatory, Karl-Schwarzschild-Str.
   2, 85748 Garching, Germany\label{inst5}
   \and
   Universit\"ats-Sternwarte M\"unchen, Scheinerstr. 1, 81679
   M\"unchen, Germany\label{inst6}
   \and
   Department of Physics, Denys Wilkinson Building, Keble Road,
   Oxford, OX1 3RH, United Kingdom\label{inst7}
   \and
   Institut f\"ur Physik und Astronomie, Universit\"at Potsdam,
   Karl-Liebknecht-Str. 24/25, 14476 Potsdam, Germany\label{inst8}
   \and
   Institut d’Astrophysique et de G\'eophysique, Universit\'e de
   Li\`ege, All\'ee du 6 Ao\^ut, B\^at. B5c, 4000 Li\`ege, Belgium
   \label{inst9}
   \and
   Anton Pannekoek Institute for Astronomy, University of Amsterdam,
   Science Park 904, PO Box 94249, 1090 GE, Amsterdam, The Netherlands
   \label{inst10}
   \and
   Instituut voor Sterrenkunde, KU Leuven, Celestijnenlaan 200D, 3001,
   Leuven, Belgium \label{inst11}
   }

   \date{}

% \abstract{}{}{}{}{} 
%  {} token are mandatory
 
  \abstract
   {}
   {We report the detection of a magnetic field in the helium-strong star 
   CPD\,$-57^{\circ}$\,3509 (B2\,IV), a member of the Galactic open cluster 
   NGC\,3293, and characterise the star's atmospheric and fundamental parameters.}
   {Spectropolarimetric observations with FORS2 and HARPSpol are
   analysed using two independent approaches to  
   quantify the magnetic field strength. A high-S/N FLAMES/GIRAFFE spectrum is
   analysed using a hybrid non-LTE model atmosphere technique.
   Comparison with stellar evolution models constrains the
   fundamental parameters of the star.}
   {We obtain a firm detection of a surface averaged longitudinal magnetic field
   with a maximum amplitude of about 1\,kG. Assuming a dipolar
   configuration of the magnetic field, this implies a 
   dipolar field strength larger than 3.3\,kG. 
   Moreover, the large amplitude and fast variation (within about 1 day) of the longitudinal 
   magnetic field implies that CPD\,$-57^{\circ}$\,3509 is spinning
   very fast despite its apparently slow projected rotational velocity. 
   The star should be able to support a centrifugal magnetosphere, yet the
   spectrum shows no sign of magnetically confined material; in
   particular, emission in H$\alpha$ is not observed. Apparently, the
   wind is either not strong enough for enough material to accumulate in the
   magnetosphere to become observable or, alternatively, some leakage
   process leads to loss of material from the magnetosphere. The quantitative 
   spectroscopic analysis of the star yields an effective temperature and a 
   logarithmic surface gravity of 23\,750$\pm$250\,K and 4.05$\pm$0.10,
   respectively, and a surface helium fraction of 0.28$\pm$0.02 by
   number. The surface abundances of C, N, O, Ne, S, and Ar are
   compatible with the cosmic abundance standard, whereas Mg, Al, Si, and Fe are
   depleted by about a factor of 2. This abundance pattern can be
   understood as the consequence of a fractionated stellar wind.
   CPD\,$-57^{\circ}$\,3509 is one of the
   most evolved He-strong stars known with an independent age constraint 
   due to its cluster membership.}
  {}
  

   \titlerunning{Magnetic field of CPD\,$-57^{\circ}$\,3509}
   \authorrunning{Przybilla et al.}

   \maketitle
\section{Introduction}  
Helium-strong stars (often also called He-rich stars) 
constitute the hottest and most massive chemically peculiar (CP) stars 
of the upper main sequence \citep[e.g.][]{Smith96}. They typically 
populate the temperature domain of $\sim$20\,000 -- 25\,000\,K, coinciding
with a spectral type around B2. Their helium lines are anomalously
strong for their colours, implying abundance ratios of He/H\,$\sim$\,0.5
by number, while their hydrogen lines are essentially normal, and the
metal lines show no outstanding anomalies \citep{Walborn83}.
The prototype He-strong star discussed in the literature is 
\object{$\sigma$\,Ori\,E} \citep{GrWa58}, and only several tens of members of this
rare class of stars are known to date. Few systematic
investigations of the properties of small samples of He-strong stars are found in
the literature \citep{Zboriletal97,ZbNo99,Leoneetal97,Cidaleetal07,Zboril11}.

\begin{figure*}[ht]
\centering
\includegraphics[width=.94\linewidth,clip]{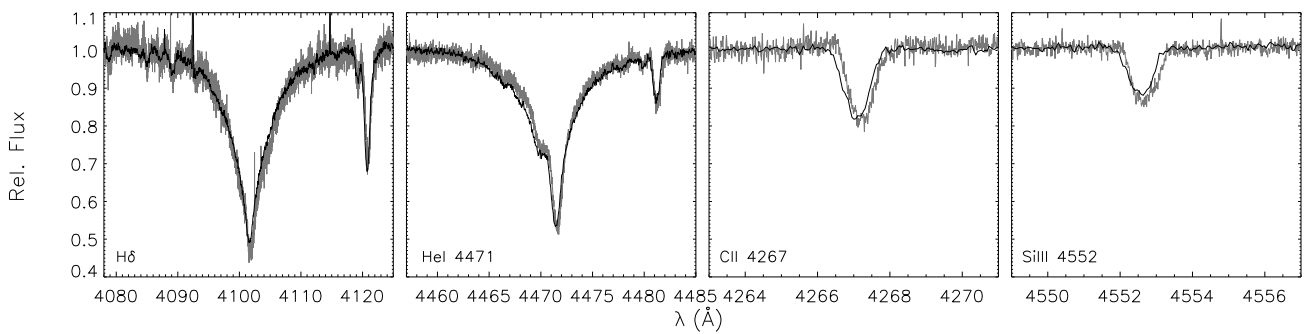}
\caption{Comparison of the FLAMES/GIRAFFE (black line) and the combined HARPSpol Stokes $I$
spectrum (grey line) of CPD\,$-57^{\circ}$\,3509 for several strategic
lines, as indicated. The observed spectra were 
cross-correlated with the model spectrum computed for our final set of parameters to determine the barycentric radial velocity, and shifted to the
laboratory frame for the comparison. See the text for~further~discussion.}
\label{fig:comparison}
\end{figure*}

The He-strong stars include some of the strong\-est magnetic 
fields detected in non-degenerate stars
\citep[e.g.][]{BoLa79,Bohlenderetal87,Hubrigetal15b}.
These magnetic fields suppress atmospheric turbulence which
allows atmospheric inhomogeneities (spots, abundance
stratifications) to develop. This takes place in the presence of weak,
fractionated stellar winds. \citet{SpPa92} demonstrated that the
assumption of a one-component fluid can break down for low-density winds
such as those in early B dwarfs. In such winds the metal ions
can lose their dynamical coupling to the ions of hydrogen and helium.
The metal ions move with high velocities, while hydrogen and helium
are not dragged along. In the context of He-strong stars, it is
particularly important that at temperatures $<$25\,000\,K helium is found
increasingly in the neutral stage. This allows helium to effectively decouple 
from the radiatively driven outflow and the magnetic confinement and fall back 
to the surface of the star, creating the observed overabundance
\citep{HuGr99,KrKu01}. The lower temperature boundary for this phenomenon 
occurs when neutral hydrogen also decouples from the outflow. Some material may also be trapped 
in a centrifugal magnetosphere \citep[see e.g.][]{petit2013,Hubrigetal15b}, 
giving rise to `double-horned' line profiles in H$\alpha$ and some near-IR hydrogen lines 
that are characteristic of the $\sigma$ Ori E analogues.

The He-strong nature of CPD\,$-57^{\circ}$\,3509 
was first noticed by \citet[designated as NGC\,3293-034]{Evansetal05} 
in a systematic spectroscopic
survey of the massive star content of the open cluster
\object{NGC\,3293}. Here we report on the first spectropolarimetric observations 
of the star (Sect.~\ref{sect:observations}) obtained within our
``B-Fields in OB Stars'' (BOB) collaboration
\citep{Moreletal14,Moreletal15} and on the magnetic field
detection (Sect.~\ref{sect:detection}). In the second part of the
paper, results from the first quantitative spectral analysis that accounts 
for the CP nature of the star are presented (Sect.~\ref{sect:models}).
The discussion of the findings concludes the present work
(Sect.~\ref{sect:discussion}).

\section{Observations\label{sect:observations}}  
We observed CPD\,$-57^{\circ}$\,3509 using the FORS2 low-resolution 
spectropolarimeter \citep{app1992,app1998} attached to the Cassegrain 
focus of the 8-m Antu telescope of the ESO Very Large Telescope of 
the Paranal Observatory. 
The observations were performed using the 2k$\times$4k MIT CCDs 
during the first visitor run in 2014 February and with the
2k$\times$4k EEV CCDs during runs in 2014 June and 2015 March. 
For all spectropolarimetric observations, we used the grism 600B,
which has an average spectral dispersion of 0.75\,{\AA}/pixel. The use
of the mosaic detector with a pixel size of 15\,$\mu$m allowed us to cover the
spectral range from 3250 to 6215\,{\AA}, which includes all
Balmer lines except H$\alpha$ and a number of He lines. 
A slit width of $0.4\,\arcsec$ was used, resulting
in a spectral resolving power of $\sim$\,1700, as measured from 
emission lines of the wavelength calibration lamp.
The star was observed once 
each night on  2014 February 6 and 7, on  2014 
June 1 and 2, and on  2015 March 17 
with a sequence of spectra obtained alternatively rotating the quarter waveplate 
from $-$45$^{\circ}$ to $+$45$^{\circ}$ every second exposure 
(i.e. $-$45$^{\circ}$, $+$45$^{\circ}$, $+$45$^{\circ}$, $-$45$^{\circ}$, 
$-$45$^{\circ}$, $+$45$^{\circ}$, etc.). The adopted exposure times and obtained 
signal-to-noise ratio (S/N) of Stokes $I$ 
are listed in Table~\ref{tab:fors}.

We observed CPD\,$-57^{\circ}$\,3509 further using the HARPSpol polarimeter
\citep{snik2011,piskunov2011} feeding the HARPS spectrograph 
\citep{mayor2003} attached to the ESO 3.6-m telescope in La\,Silla, Chile. 
The observations, covering the 3780--6910\,\AA\ wavelength range with a 
spectral resolution $R\sim$115\,000 were obtained on the 23 April 2014 
using the circular polarisation analyser. We observed the star with a 
sequence of four sub-exposures obtained rotating the quarter-wave retarder 
plate by 90$^\circ$ after each exposure, i.e. 45$^\circ$, 135$^\circ$, 
225$^\circ$, and 315$^\circ$. Exposure times per sub-exposure were 2700\,s, 
leading to a Stokes $I$ S/N per pixel, 
calculated at $\lambda$4950\,\AA, of $\sim$100.

Finally, an optical spectrum covering the wavelength range
3850--4755\,{\AA} and 6380--6610\,{\AA} with S/N\,$\approx$\,250 (at about
4725\,{\AA}) and $R$ varying between $\sim$20\,000-30\,000 is at our disposal, 
taken on 26/02/2000 with FLAMES/GIRAFFE on the ESO VLT \citep{Pasquinietal00}. 
See \citet{Evansetal05} and \citet{Maederetal14} for details 
of the observations, data reduction, and post-processing of the spectra. 
An overcorrection for sky emission in H$\alpha$ was detected
in the data and corrected~here.

A comparison of the FLAMES/GIRAFFE and the HARPS\-pol Stokes $I$
spectrum from the coaddition of the four exposures is shown in Fig.~\ref{fig:comparison}
for H$\delta$, a strong \ion{He}{i} line, and two metal lines. 
The two spectra were cross-correlated with the model spectrum computed for our final set of
parameters (Sect.~\ref{sect:results}) in order to determine the
respective barycentric radial velocities, and shifted to
the laboratory frame for the comparison. The small line
asymmetries and shifts in the central wavelengths of the lines (by
$\sim$4\,km\,s$^{-1}$) are compatible with the presence of spots 
on the surface, which are common in He-strong stars and cannot be
taken as evidence for binarity. However, overall the line strengths in both
spectra are rather similar. Therefore, the FLAMES/GIRAFFE spectrum was adopted 
for the model atmosphere analysis because of its higher S/N.
A further comparison with the four individual HARPSpol exposures with
the FLAMES/GIRAFFE spectrum is omitted here because of their low S/N.

\begin{table*}[th!]
\caption[ ]{Average longitudinal magnetic field values obtained from
the FORS2 observations.\\[-9mm]}
\label{tab:fors}
\begin{center}
\begin{footnotesize}
\begin{tabular}{l|cc|ccc|r@{$\,\pm\,$}lr@{$\,\pm\,$}lr@{$\,\pm\,$}lr@{$\,\pm\,$}l}
\hline
\hline
Reduction & Date & HJD$-$  & \# of  & Exp. & S/N & \multicolumn{2}{c}{\bz}
& \multicolumn{2}{c}{\nz}          & \multicolumn{2}{c}{\bz}    & \multicolumn{2}{c}{\nz}          \\
          &      & 2450000 & frames & time &     &        \multicolumn{4}{c}{Hydrogen} & \multicolumn{4}{c}{All} \\
\hline
Bonn    & 06/02/2014 & 6695.72851 & 8 & 344 & 1380 & $-$356&125 & $-$361&126 & $-$143&78 & $-$39&78 \\
        & 07/02/2014 & 6696.77037 & 8 & 327 & 1828 &    659&109 & $-$120&97  &    710&58 &    68&56 \\
        & 01/06/2014 & 6810.48164 & 8 & 550 & 1943 &  $-$71&71  &  $-$58&77  &     40&46 & $-$51&47 \\
        & 02/06/2014 & 6811.46357 & 8 & 550 & 2289 &   1049&69  &  $-$90&61  &    943&41 &     1&39 \\
        & 17/03/2015 & 7099.57011 & 8 & 562 & 1791 &    607&98  &   $-$4&89  &    734&56 &     8&55 \\
\hline
Potsdam & 06/02/2014 & 6695.72851 & 8 & 344 & 1381 & $-$287&126 & $-$377&139 &  $-$23&60 & $-$101&64 \\
        & 07/02/2014 & 6696.77037 & 8 & 327 & 1826 &    694&108 & $-$116&104 &    539&51 &      1&48 \\
        & 01/06/2014 & 6810.48164 & 8 & 550 & 2025 &  $-$19&71  & $-$28&86   &     88&54 &  $-$45&59 \\
        & 02/06/2014 & 6811.46357 & 8 & 550 & 2348 &    979&68  & $-$108&77  &    920&48 &      2&50 \\
        & 17/03/2015 & 7099.57011 & 8 & 562 & 1826 &    582&99  & $-$75&101  &    671&62 &  $-$33&61 \\
\hline
\end{tabular}
\end{footnotesize}
\end{center}
\vspace{-5mm}
\end{table*}

\section{Magnetic field detection\label{sect:detection}}
\subsection{FORS2 observations}
Because of several controversies present in the literature about 
magnetic field detections and measurements performed with the FORS 
spectropolarimeters \citep[see e.g.][]{bagnulo2012,Hubrigetal15a}, the data have been 
independently reduced by two different groups (one in Bonn and one in Potsdam), 
each using different and completely independent tools and routines. 
The first reduction and analysis (Bonn) was performed employing 
IRAF\footnote{Image Reduction and Analysis Facility 
(IRAF -- {\tt http://iraf.noao.edu/}) is distributed by the National 
Optical Astronomy Observatory, which is operated by the Association of 
Universities for Research in Astronomy (AURA) under cooperative agreement 
with the National Science Foundation.} \citep{tody} and IDL routines based 
on the technique and recipes presented by \citet{bagnulo2002,bagnulo2012}. 
The second reduction and analysis (Potsdam) was based on the tools described 
in \citet{hubrig2004a,hubrig2004b}. Details of the data reduction and analysis 
procedure applied by both groups are given in a separate dedicated 
paper \citep{Fossatietal15b}.

%--------------------------------------------------------------------
\begin{figure}[t]
\centering
\includegraphics[width=86mm,clip]{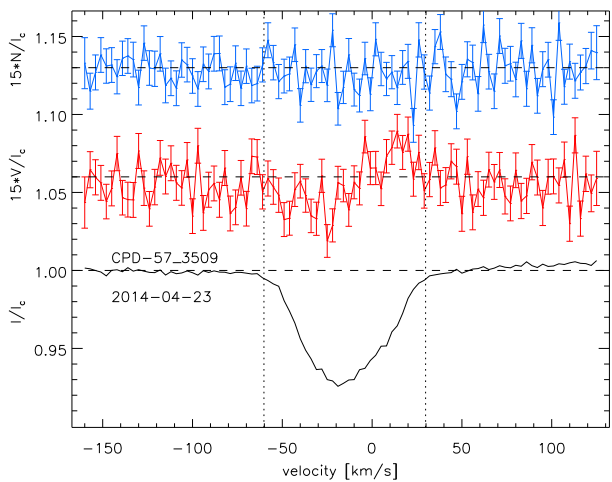}
\caption{LSD profiles of Stokes $I$ (black solid line -- bottom profile), 
$V$ (red solid line -- middle profile), and $N$ parameter (blue solid line 
-- top profile) obtained for CPD\,$-57^{\circ}$\,3509 in the Bonn
reduction. The error bars of the LSD 
profiles are shown for both Stokes $V$ and the $N$ parameter. The 
vertical dotted lines indicate the velocity range adopted for the 
determination of the detection probability and magnetic field value. 
All profiles have been rigidly shifted upwards/downwards using arbitrary 
values and the Stokes $V$ and $N$ profiles have been expanded 15 times.}
\label{fig:lsd}
\end{figure}
%--------------------------------------------------------------------
%

The value of the surface-averaged longitudinal magnetic field 
\bz\ was calculated using either the hydrogen lines or the whole 
spectrum in the 3710--5870\,\AA\ spectral region. A summary of the
results is shown in Table~\ref{tab:fors}. 
The first column indicates the group, hence adopted data
reduction and analysis, which obtained the results shown in the
remaining columns. The heliocentric Julian date shown in Column 3
is that of the beginning of the sequence of exposures. Column 4
gives the number of frames obtained during each night of observation,
while column 5 shows the exposure time, in seconds, of each frame.
Column 6 gives the S/N per pixel of Stokes $I$ calculated at about
4950\,\AA\ over a wavelength range of 100\,\AA. 
Columns 7 and 8 give the \bz\ values in Gauss, obtained using the 
spectral regions covered by the hydrogen lines obtained from the Stokes 
$V$ and $N$ parameter spectrum, respectively. The same is given in Columns 
9 and 10, but using the entire spectrum.

Of the five FORS2 
measurements, those obtained on the 7 February 2014, 2 June 2014, 
and 17 March 2015 led to a magnetic field detection, while we 
consistently found non-detections from the null profile (i.e. 
\nz\ consistent with zero). The results from both independent data
reductions and analyses are consistent within the mutual uncertainties.

\subsection{HARPS observations}
Like for the FORS2 data, the HARPSpol observations have been 
independently reduced by two different groups. 
The first reduction and analysis (Bonn) was performed with the 
{\sc reduce} package \citep{piskunov2002} and the least-squares 
deconvolution technique \citep[LSD;][]{donati1997}, while the second 
reduction and analysis (Potsdam) was performed with the ESO/HARPS 
pipeline and the LSD technique. 

\subsubsection{Bonn reduction and analysis}
The one-dimensional spectra, obtained with {\sc reduce}, were combined 
using the ``ratio'' method in the way described by \citet{bagnulo2009}. 
We then re-normalised all spectra to the intensity of the continuum 
obtaining a spectrum of Stokes $I$ ($I/I_c$) and $V$ ($V/I_c$), plus 
a spectrum of the diagnostic null profile \citep[$N$ - see][]{bagnulo2009}, 
with the corresponding uncertainties. The profiles of the 
Stokes $I$, $V$, and $N$ parameter were analysed using LSD, which combines line 
profiles (assumed to be identical) centred on the position of the 
individual lines and scaled according to the line strength and 
sensitivity to a magnetic field (i.e. line wavelength and Land{\'e} 
factor). We computed the LSD profiles of Stokes $I$, $V$ and of the 
null profile using the methodology and the code described in 
\citet{kochukhov2010}. We prepared the line mask used by the LSD code 
adopting the stellar parameters obtained from the spectroscopic analysis 
(see Sect.~\ref{sect:models}). We extracted the line parameters from the Vienna 
Atomic Line Database \citep[{\sc vald};][]{vald1,vald2,vald3} and tuned 
the given line strength to the observed Stokes $I$ spectrum with the aid 
of synthetic spectra calculated with {\sc synth3} \citep{kochukhov2007}. 
We used all lines deeper than 10\% of the continuum, avoiding hydrogen 
lines, helium lines with developed wings, and lines in spectral regions 
affected by the presence of telluric features,
91 lines in total. We defined the magnetic 
field detection making use of the false alarm probability 
\citep[FAP;][]{donati1992}, considering a profile with 
FAP\,$<$\,10$^{-5}$ as a definite detection, 
$10^{-5}$\,$<$\,FAP\,$<$\,$10^{-3}$ as a marginal detection, 
and FAP\,$>$\,$10^{-3}$ as a non-detection. 

The LSD profiles we obtained for CPD\,$-57^{\circ}$\,3509 are shown in 
Fig.~\ref{fig:lsd}, with the S/N of the LSD Stokes $V$ profile reaching 
1342. The analysis of the Stokes $V$ LSD profile led to a clear, 
definite detection with a FAP\,=\,9.4$\times$10$^{-7}$, while the 
analysis of the LSD profile of the null parameter led to a
non-detection, $\left<N_{\rm z}\right>$\,=\,75$\pm$72\,G with
FAP\,$>$\,0.01. Integrating over a range of 90\,$\mathrm{km\,s}^{-1}$ 
(i.e. $\pm$45\,$\mathrm{km\,s}^{-1}$ from the line centre), we 
derived \bz\,=\,$-$557$\pm$73\,G. 
The measurements carried out with FORS2 and HARPS indicate the 
presence of a rather strong field with reversing polarity.

\begin{figure}[t]
\centering
\includegraphics[width=77mm,height=90mm,clip]{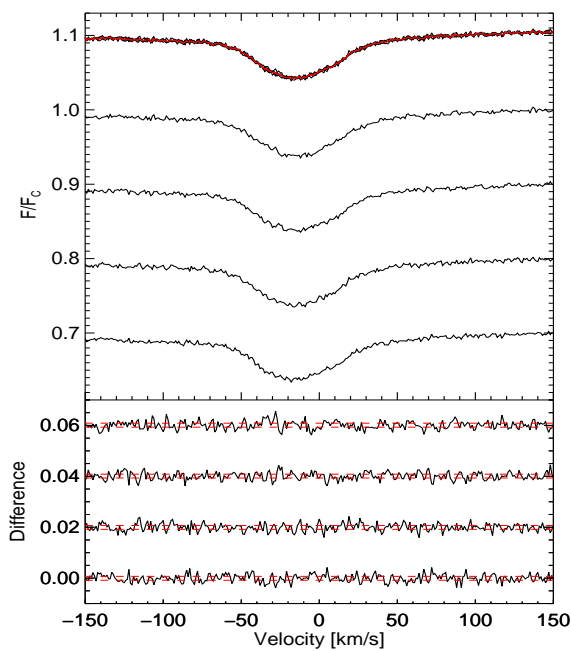}
\caption{Testing for the presence of $\beta$ Cephei pulsations in
CPD\,$-57^{\circ}$\,3509 on the basis of LSD Stokes $I$ profiles for the 
individual HARPS subexposures from the Potsdam reduction.
The normalised average profile (in red, overlaid on all four
sub-exposures in the topmost curve) and the individual subexposure profiles 
(subsequent curves) are shown in the upper
panel, shifted by constant offsets. The lower panel shows the difference 
between the individual subexposures and the average profile, with the
dashed lines indicating the $\pm$1$\sigma$-ranges.}
\label{fig:pulsations}
\end{figure}

\subsubsection{Potsdam reduction and analysis}
The reduction and calibration of the HARPS polarimetric spectra were performed
using the HARPS data reduction software available at ESO's 3.6\,m telescope
(La~Silla, Chile). The normalisation of the spectra to the continuum
level consisted of several steps described in detail by \citet{Hubrigetal13}.
The Stokes~$I$ and $V$ parameters were derived following 
\citet{Ilyin12}, and null polarisation spectra were 
calculated by combining the subexposures in such a way that polarisation canceled out.
These steps minimise spurious signals in the obtained data \citep[e.g.][]{Ilyin12}.

Since a number of magnetic Bp stars were reported to show $\beta$~Cephei-like
pulsations \citep[e.g.][]{Neineretal12}, and pulsations are known to have an impact
on the analysis of the presence of a magnetic field and its strength
\citep[e.g.][]{Schnerretal06,Hubrigetal11a}, we verified that no change
in the line profile shape or radial velocity shifts are present in the obtained 
sub-exposures. We recall that the time elapsed between
consecutive exposures is 45 minutes. On the one hand, this time scale is appropriate 
to detect $\beta$\,Cep-like pulsations \citep[which typically have periods of
4 hours, e.g.][]{StHa05}. On the other hand, the total time span 
(3 hours) is likely to be short enough to be free
of the effects of the rotational modulation by spots.
In Fig.~\ref{fig:pulsations} we present Stokes~$I$ profiles computed
for the individual subexposures. The line mask consisting of 163 \ion{He}{i} and 
metal lines was constructed using the VALD database. The overplotted profiles are shown on the top,
together with the average profile.
The differences between the Stokes~$I$ profiles computed for the individual 
sub-exposures and the average Stokes~$I$ profile are presented in the lower panel.
No impact of pulsations at a level higher than the spectral noise is
detected, which is in line with the non-detection of photometric
variations by \citet{Balona94} in his search for short period B-type
variables in NGC\,3293.

\begin{figure}[t]
\centering
\includegraphics[width=78mm,height=90mm,clip]{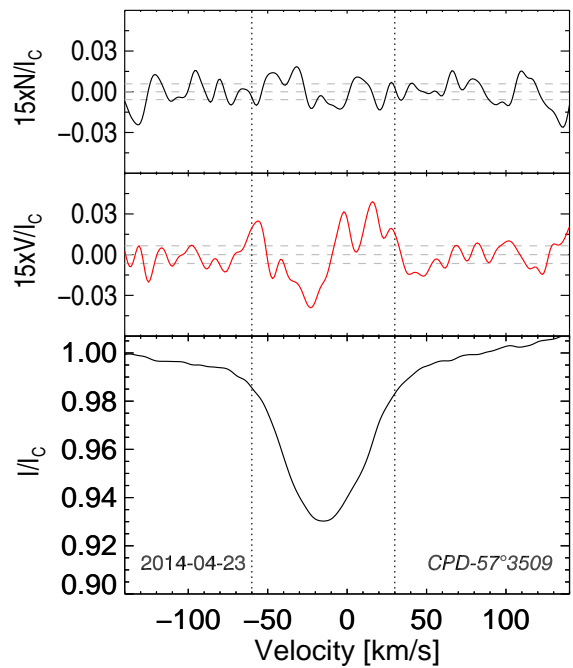}
\caption{Same as Fig.~\ref{fig:lsd}, but for the Potsdam reduction.
The horizontal dashed lines indicate the average values and the
$\pm$1$\sigma$-ranges.}
\label{fig:lsd2}
\end{figure}

To search for the magnetic field, we employed an independent
implementation of the LSD technique once more. The resulting mean LSD Stokes~$I$, Stokes~$V$, and
diagnostic $N$ profiles obtained for the same line list as used for
the search of the spectral variability are presented in
Fig.~\ref{fig:lsd2}. 
Using the FAP in the region of the whole
Stokes~$I$ line profile (velocity range from $-$60\,km\,s$^{-1}$ to
+30\,km\,s$^{-1}$), we obtained a definite magnetic field detection 
$\left<B_{\rm z}\right>$\,=\,$-$490$\pm$29\,G
with FAP\,=\,$1\times 10^{-10}$. The null parameter led to a
non-detection, $\left<N_{\rm z}\right>$\,=\,34$\pm$50\,G with FAP\,$=$\,0.024.
We conclude that the two independent LSD analyses are consistent.

\section{Model atmosphere analysis\label{sect:models}}
\subsection{Codes and analysis methodology}
We employ a hybrid non-LTE approach for the model atmosphere analysis
of CPD\,$-57^{\circ}$\,3509 
\citep[henceforth abbreviated as NP12]{NiPr07,NiPr12}. Model atmospheres were
computed with the {\sc Atlas9} code \citep{Kurucz93}, which assumes 
plane-parallel geometry, chemical
homogeneity, and hydrostatic, radiative, and local
thermodynamic equilibrium (LTE). The model atmospheres were held
fixed in the subsequent non-LTE line-formation calculations. 
Non-LTE level populations
and model spectra were obtained with recent versions of {\sc Detail}
and {\sc Surface} \citep[both updated and extended by one of us (KB)]{Giddings81,BuGi85}. 
The coupled radiative transfer and statistical
equilibrium equations were solved with {\sc Detail}, employing
the accelerated lambda iteration scheme of \citet{RyHu91}. 
This allowed even complex ions to be treated in a realistic
way. Synthetic spectra were calculated with {\sc Surface},
using refined line-broadening theories. Continuous opacities
due to hydrogen and helium were considered in non-LTE, and
line blocking was accounted for via LTE opacity sampling, employing
the method of \citet{Kurucz96}.
Microturbulence was considered in a consistent way throughout
all computation steps: atmospheric structure computations, non-LTE level 
populations determination, and the formal solution.

The He-strong nature of CPD\,$-57^{\circ}$\,3509 precludes the use of existing model
grids for chemically normal stars for the analysis.
Instead, dedicated computations were required to constrain the 
atmospheric parameters and elemental abundances.
Non-LTE level populations and the synthetic spectra of all elements
were calculated using the suite of model atoms listed
in Table~\ref{atoms}. Updates of some of the published models were carried
out by introducing improved oscillator strengths and collisional
data. These model atoms were employed previously by NP12 for the 
analysis of a sample of normal early B-type stars in a parameter range
similar to that expected for CPD\,$-57^{\circ}$\,3509. They were
complemented by model atoms for additional trace species, which facilitated 
the determination of abundances for all elements detectable in the 
available high-resolution spectrum.

\begin{table}[t!]
\footnotesize
\caption[]{Model atoms for non-LTE calculations.\\[-6mm] \label{atoms}}
\setlength{\tabcolsep}{.15cm}
\begin{tabular}{ll}
\hline
\hline
\footnotesize
Ion     &  Model atom \\
\hline\\[-3mm]
H               &  \citet{PrBu04}\\
\ion{He}{i/ii}  &  \citet{Przybilla05}\\
\ion{C}{ii/iii} &  \citet{NiPr06,NiPr08}\\
\ion{N}{ii}     &  \citet{PrBu01}, updated\tablefootmark{a}\\
\ion{O}{ii}     &  \citet{BeBu88}, updated\tablefootmark{a}\\
\ion{Ne}{i}     &  \citet{MoBu08}, updated\tablefootmark{a}\\
\ion{Mg}{ii}    &  \citet{Przybillaetal01a}\\
\ion{Al}{iii}   &  Przybilla (in prep.)\\
\ion{Si}{ii/iii/iv} &  Przybilla \& Butler (in prep.)\\
\ion{S}{ii/iii} &  \citet{Vranckenetal96}, updated\tablefootmark{a}\\
\ion{Ar}{ii}    &  Butler (in prep.)\\
\ion{Fe}{iii}   &  \citet{Moreletal06}, corrected\tablefootmark{b}\\
\hline\\[-5mm]
\end{tabular}
\tablefoot{
\tablefootmark{a}{See
Table~\ref{tab:linelist} for details.}
\tablefootmark{b}{See NP12.}}
\end{table}

In a first step, the hydrogen Balmer lines, the \ion{He}{i/ii} lines,
and additional ionization equilibria\footnote{Ionisation
equilibria are established when the abundances of the different
ionisation stages of an element agree within the uncertainties
for a given set of atmospheric parameters.} of \ion{C}{ii/iii},
\ion{Si}{ii/iii/iv} and \ion{S}{ii/iii} were employed to constrain the 
atmospheric parameters -- effective temperature $T_\mathrm{eff}$,
(logarithmic) surface gravity $\log g$, microturbulence $\xi$, helium
abundance $y$ (by number), projected rotational velocity $v \sin i$ and 
macroturbulent velocity $\zeta$ -- in a similar approach to NP12. 
We use {\sc Spas} \citep[Spectrum Plotting and Analysing
Suite,][]{Hirsch09} for the comparison of synthetic spectra with
observations based on microgrids. {\sc Spas} provides the means to interpolate between
model grid points for up to three parameters simultaneously
and allows instrumental, rotational, and (radial-tangential)
macrobroadening functions to be applied to the resulting theoretical
profiles. The program uses the downhill simplex algorithm
\citep{NeMe65} to minimise $\chi^2$ in order to find a good fit
to the observed spectrum. Once the atmospheric parameters were established,
elemental abundances for the additional chemical species were
determined using {\sc Spas}. With the resulting abundances the entire process was 
iterated to account for the abundance effects on line
blanketing and blocking.\\[-8mm]
\paragraph{Limitations.} The present approach does not account for phenomena like spots or
vertical chemical stratification of the atmosphere, or for the
effects of the magnetic field on the radiative transfer. Such effects
are modelled occasionally using LTE techniques for spectrum synthesis
on prescribed model atmospheres
\citep[e.g.][]{Landstreet88,Donati01,Carrolletal12}, but 
non-LTE effects -- which are important in early B-type stars
\citep[e.g.][]{NiPr07,Przybillaetal11} -- are just being
considered \citep{Yakuninetal15}.
While the two available high-resolution spectra indicate the
presence of surface spots because of the small-scale line-profile
changes\footnote{The presence of spots can influence line
profiles in various ways, giving rise to e.g. (periodic) line asymmetries, shifts
in the line centroids, and/or changes in equivalent widths, which are tied to the
stellar rotation. Different chemical elements may show different distributions over 
the stellar surface, see e.g. the Doppler imaging work on the prototype He-strong star
$\sigma$\,Ori E by \citet{Oksalaetal15}.} (see Fig.~\ref{fig:comparison} for examples), 
the deviations from a homogeneous 
surface and resulting symmetric line profiles generally seem small, as implied by
the good fit of the model to the FLAMES/GIRAFFE spectral
snapshot (see Sect.~\ref{sect:results}). The issue may be revisited 
in greater detail once proper high-quality time series observations 
of the star become available.

Other effects that are not considered are the potential oblateness and
gravity darkening in a fast-spinning star. Some He-strong
stars are known to rotate near critical velocity
\citep{Grunhutetal12,Riviniusetal13}, and the nightly variation in the
magnetic field (Sect.~\ref{sect:detection}) implies that
CPD\,$-57^{\circ}$\,3509 is rotating at a much higher velocity than the 
observed rather low $v \sin i$\,=\,35\,km\,s$^{-1}$ would suggest. The question is 
how close to critical velocity ($\sim$500\,km\,s$^{-1}$ in this case) 
the star rotates, since significant effects on the atmospheric
parameter and abundance determination are expected only for rotational
velocities over 60\% critical \citep{Frematetal05}. 
Assuming that CPD\,$-57^{\circ}$\,3509 is a magnetic oblique rotator, the observed 
change in polarity of the magnetic field implies that
the equatorial rotational velocity should be in the range $\sim$70--250\,km\,s$^{-1}$ 
assuming a one to three-day rotation period (see 
Sect.~\ref{sect:discussion} for a discussion), so our present analysis 
approach seems adequate. However, a conclusive statement on
this can only be given once the rotation period is firmly established.

\begin{table}[t]
\caption[ ]{Parameters and elemental abundances of
CPD\,$-57^{\circ}$\,3509.\\[-9mm]}
\label{tab:parameters}
\begin{center}
\begin{footnotesize}
\begin{tabular}{lll}
\hline
\hline
Sp.~Type                   & B2\,IV He-strong\\
$v_\mathrm{rad}$           & $-$16 {\ldots} $-20$\,km\,s$^{-1}$\\
$d_\mathrm{spec}$          & 2630\,$\pm$\,370\,pc\\
$B_\mathrm{d}$             & $\gtrsim$3300\,G\\[1.3mm]
\multicolumn{3}{l}{Atmospheric parameters:}\\
$T_\mathrm{eff}$           & 23750\,$\pm$\,250\,K\\
$\log g$\,(cgs)            & 4.05\,$\pm$\,0.10\\
$y$\,(number fraction)     & 0.28\,$\pm$\,0.02\\
$\xi$                      & 2\,$\pm$\,1\,km\,s$^{-1}$\\
$v \sin i$                 & 35\,$\pm$\,2\,km\,s$^{-1}$\\
$\zeta$                    & 10\,$\pm$\,2\,km\,s$^{-1}$\\[1.3mm]
\multicolumn{3}{l}{Non-LTE metal abundances:}\\
$\log$(C/H)\,$+$\,12       & 8.37\,$\pm$\,0.09\,(5)\\
$\log$(N/H)\,$+$\,12       & 7.70\,$\pm$\,0.07\,(20)\\
$\log$(O/H)\,$+$\,12       & 8.65\,$\pm$\,0.08\,(36)\\
$\log$(Ne/H)\,$+$\,12      & 8.05\,$\pm$0.15\,(2)\\
$\log$(Mg/H)\,$+$\,12      & 7.17\,(1)\\
$\log$(Al/H)\,$+$\,12      & 5.93\,$\pm$\,0.07\,(4)\\
$\log$(Si/H)\,$+$\,12      & 7.16\,$\pm$0.05\,(7)\\
$\log$(S/H)\,$+$\,12       & 7.17\,$\pm$\,0.04\,(4)\\
$\log$(Ar/H)\,$+$\,12      & 6.68\,$\pm$\,0.04\,(2)\\
$\log$(Fe/H)\,$+$\,12      & 7.30\,$\pm$\,0.04\,(6)\\[1.3mm]
\multicolumn{3}{l}{Photometric data:}\\
$V$                        & 10.68\,$\pm$\,0.06\,mag\\
$B-V$                      & ~~0.10\,$\pm$\,0.03\,mag\\
$E(B-V)$                   & ~~0.33\,$\pm$\,0.03\,mag\\
$M_V$                      & $-$2.47\,$\pm$\,0.33\,mag\\
$M_\mathrm{bol}$           & $-$4.86\,$\pm$\,0.33\,mag\\[1.3mm]
\multicolumn{3}{l}{Fundamental parameters:}\\
                           & \citet{Ekstroemetal12} & \citet{Brottetal11}\\
                           & tracks                 & tracks/{\sc Bonnsai}\\
$M/M_\odot$                & 9.7$\pm$0.3 & 9.2$\pm$0.4\\
$R/R_\odot$                & 5.0$\pm$0.9 & 4.4$^{+0.7}_{-0.5}$\\
$\log L/L_\odot$           & 3.85$\pm$0.13 & 3.76$^{+0.12}_{-0.11}$\\
$\tau$                     & 13.8$^{+2.4}_{-3.3}$\,Myr & 13.0$^{+1.7}_{-4.0}$\,Myr\\
$\tau/\tau_\mathrm{MS}$    & 0.51$^{+0.12}_{-0.16}$ & 0.47$^{+0.13}_{-0.17}$\\[.5mm]
\hline\\[-5mm]
\end{tabular}
\tablefoot{1$\sigma$-uncertainties are given. For abundances these are
from the line-to-line scatter, systematic errors amount to an
additional $\sim$0.1\,dex.\\[-1cm]}
\end{footnotesize}
\end{center}
\end{table}
%
%

%should appear online only
\begin{longtab}
\begin{longtable}{l@{\hspace{.7mm}}lrrlllr}
\caption{Spectral line analysis for CPD\,$-57^{\circ}$\,3509\label{tab:linelist}}\\[-2mm]
\hline\hline
Ion           & $\lambda\,$(\AA) & $\chi$\,(eV) & $\log gf$ & Acc. & Src. & Broad. & $\varepsilon_\mathrm{NLTE}(X)$\\
\hline
\endfirsthead
\caption{continued.}\\[-2mm]
\hline\hline
Ion           & $\lambda\,$(\AA) & $\chi$\,(eV) & $\log gf$ & Acc. & Src. & Broad. & $\varepsilon_\mathrm{NLTE}(X)$\\
\hline
\endhead
\hline
\endfoot
\ion{C}{ii}   & 3920.69 & 16.33 & $-$0.232 & B  & WFD & C   & 8.32\\
\ion{C}{ii}   & 4267.00 & 18.05 &    0.563 & C+ & WFD & G   & 8.28\\
\ion{C}{ii}   & 4267.26 & 18.05 &    0.716 & C+ & WFD & G\\
\ion{C}{ii}   & 4267.26 & 18.05 & $-$0.584 & C+ & WFD & G\\
\ion{C}{ii}   & 6578.05 & 14.45 & $-$0.087 & C+ & N02 & C   & 8.31\\
\ion{C}{ii}   & 6582.88 & 14.45 & $-$0.388 & C+ & N02 & C   & 8.48\\[.9mm]
\ion{C}{iii}  & 4647.42 & 29.53 &    0.070 & B+ & WFD & C   & 8.45\\[.9mm]
\ion{N}{ii}   & 3955.85 & 21.15 & $-$0.813 & B  & WFD & C  & 7.75\\
\ion{N}{ii}   & 3995.00 & 18.50 &    0.163 & B  & FFT & C  & 7.60\\
\ion{N}{ii}   & 4035.08 & 23.12 &    0.599 & B  &BB89 & C  & 7.79\\
\ion{N}{ii}   & 4041.31 & 23.14 &    0.748 & B  & MAR & C  & 7.70\\
\ion{N}{ii}   & 4043.53 & 23.13 &    0.440 & C  & MAR & C  & 7.73\\
\ion{N}{ii}   & 4176.16 & 23.20 &    0.316 & B  & MAR & C  & 7.72\\
\ion{N}{ii}   & 4179.67 & 23.25 & $-$0.090 & X  & KB  & C  & 7.81\\
\ion{N}{ii}   & 4227.74 & 21.60 & $-$0.060 & B  & WFD & G  & 7.71\\
\ion{N}{ii}   & 4236.91 & 23.24 &    0.383 & X  & KB  & C  & 7.59\\
\ion{N}{ii}   & 4237.05 & 23.24 &    0.553 & X  & KB  & C  & \\
\ion{N}{ii}   & 4241.24 & 23.24 & $-$0.337 & X  & KB  & C  & 7.68\\
\ion{N}{ii}   & 4241.76 & 23.24 &    0.210 & X  & KB  & C  & \\
\ion{N}{ii}   & 4241.79 & 23.25 &    0.713 & X  & KB  & C  & \\
\ion{N}{ii}   & 4242.50 & 23.25 & $-$0.337 & X  & KB  & C  & \\
\ion{N}{ii}   & 4432.74 & 23.42 &    0.580 & X  & KB  & C  & 7.63\\
\ion{N}{ii}   & 4447.03 & 20.41 &    0.221 & B  & FFT & C  & 7.70\\
\ion{N}{ii}   & 4530.41 & 23.47 &    0.604 & C+ & MAR & C  & 7.65\\
\ion{N}{ii}   & 4601.48 & 18.47 & $-$0.452 & B+ & FFT & C  & 7.75\\
\ion{N}{ii}   & 4607.15 & 18.46 & $-$0.522 & B+ & FFT & C  & 7.69\\
\ion{N}{ii}   & 4613.87 & 18.47 & $-$0.622 & B+ & FFT & C  & 7.80\\
\ion{N}{ii}   & 4621.39 & 18.47 & $-$0.538 & B+ & FFT & C  & 7.67\\
\ion{N}{ii}   & 4630.54 & 18.48 &    0.080 & B+ & FFT & C  & 7.62\\
\ion{N}{ii}   & 4643.08 & 18.48 & $-$0.371 & B+ & FFT & C  & 7.62\\
\ion{N}{ii}   & 4694.64 & 23.57 &    0.100 & X  & KB  & C  & 7.72\\[.9mm]
\ion{O}{ii}   & 3911.96 & 25.66 & $-$0.014 & B+ & FFT & C  & 8.59\\
\ion{O}{ii}   & 3912.12 & 25.66 & $-$0.907 & B+ & FFT & C  & \\
\ion{O}{ii}   & 3945.04 & 23.42 & $-$0.711 & B+ & FFT & C  & 8.78\\
\ion{O}{ii}   & 3954.36 & 23.42 & $-$0.402 & B+ & FFT & C  & 8.69\\
\ion{O}{ii}   & 4069.62 & 25.63 &    0.144 & B+ & FFT & C  & 8.57\\
\ion{O}{ii}   & 4069.88 & 25.64 &    0.352 & B+ & FFT & C  & \\
\ion{O}{ii}   & 4072.72 & 25.65 &    0.528 & B+ & FFT & C  & 8.74\\
\ion{O}{ii}   & 4075.86 & 25.67 &    0.693 & B+ & FFT & C  & 8.75\\
\ion{O}{ii}   & 4078.84 & 25.64 & $-$0.287 & B+ & FFT & C  & 8.73\\
\ion{O}{ii}   & 4129.32 & 25.84 & $-$0.943 & B+ & FFT & C  & 8.73\\
\ion{O}{ii}   & 4132.80 & 25.83 & $-$0.067 & B+ & FFT & C  & 8.61\\
\ion{O}{ii}   & 4156.53 & 25.85 & $-$0.706 & B+ & FFT & C  & 8.79\\
\ion{O}{ii}   & 4185.45 & 28.36 &    0.604 & D  & WFD & C  & 8.52\\
\ion{O}{ii}   & 4189.58 & 28.36 & $-$0.828 & D  & WFD & C  & 8.54\\
\ion{O}{ii}   & 4189.79 & 28.36 &    0.717 & D  & WFD & C  & \\
\ion{O}{ii}   & 4317.14 & 22.97 & $-$0.368 & B+ & FFT & C  & 8.53\\
\ion{O}{ii}   & 4319.63 & 22.98 & $-$0.372 & B+ & FFT & C  & 8.57\\
\ion{O}{ii}   & 4325.76 & 22.97 & $-$1.095 & B  & FFT & C  & 8.73\\
\ion{O}{ii}   & 4349.43 & 23.00 &    0.073 & B+ & FFT & C  & 8.67\\
\ion{O}{ii}   & 4351.26 & 25.66 &    0.202 & B+ & FFT & C  & 8.61\\
\ion{O}{ii}   & 4351.46 & 25.66 & $-$1.013 & B  & FFT & C  & \\
\ion{O}{ii}   & 4366.89 & 23.00 & $-$0.333 & B+ & FFT & C  & 8.53\\
\ion{O}{ii}   & 4369.28 & 26.23 & $-$0.383 & B+ & FFT & C  & 8.69\\
\ion{O}{ii}   & 4414.90 & 23.44 &    0.207 & B  & FFT & C  & 8.62\\
\ion{O}{ii}   & 4416.97 & 23.42 & $-$0.043 & B  & FFT & C  & 8.67\\
\ion{O}{ii}   & 4452.38 & 23.44 & $-$0.767 & B  & FFT & C  & 8.63\\
\ion{O}{ii}   & 4590.97 & 25.66 &    0.331 & B+ & FFT & C  & 8.59\\
\ion{O}{ii}   & 4595.96 & 25.66 & $-$1.022 & B  & FFT & C  & 8.58\\
\ion{O}{ii}   & 4596.18 & 25.66 &    0.180 & B+ & FFT & C  & \\
\ion{O}{ii}   & 4638.86 & 22.97 & $-$0.324 & B+ & FFT & C  & 8.68\\
\ion{O}{ii}   & 4641.81 & 22.98 &    0.066 & B+ & FFT & C  & 8.69\\
\ion{O}{ii}   & 4649.13 & 23.00 &    0.324 & B+ & FFT & C  & 8.74\\
\ion{O}{ii}   & 4661.63 & 22.98 & $-$0.269 & B+ & FFT & C  & 8.69\\
\ion{O}{ii}   & 4673.73 & 22.98 & $-$1.101 & B  & FFT & C  & 8.71\\
\ion{O}{ii}   & 4676.24 & 23.00 & $-$0.410 & B+ & FFT & C  & 8.69\\
\ion{O}{ii}   & 4696.35 & 23.00 & $-$1.377 & B  & FFT & C  & 8.62\\
\ion{O}{ii}   & 4699.01 & 28.51 &    0.418 & D  & WFD & C  & 8.60\\
\ion{O}{ii}   & 4699.22 & 26.23 &    0.238 & B+ & FFT & C  & \\
\ion{O}{ii}   & 4701.18 & 28.83 &    0.088 & C  & WFD & C  & 8.68\\
\ion{O}{ii}   & 4703.16 & 28.51 &    0.262 & D  & WFD & C  & 8.69\\
\ion{O}{ii}   & 4705.35 & 26.25 &    0.533 & B+ & FFT & C  & 8.57\\
\ion{O}{ii}   & 4710.01 & 26.23 & $-$0.090 & B+ & FFT & C  & 8.63\\[.9mm]
\ion{Ne}{i}   & 6402.25 & 16.62 &    0.365 & B+ & FFT & C  & 7.95\\
\ion{Ne}{i}   & 6506.53 & 16.67 & $-$0.002 & B+ & FFT & C  & 8.16\\[.9mm]
\ion{Mg}{ii}  & 4481.13 &  8.86 &    0.730 & B+ & FW  & G  & 7.17\\
\ion{Mg}{ii}  & 4481.15 &  8.86 & $-$0.570 & B+ & FW  & G  & \\
\ion{Mg}{ii}  & 4481.33 &  8.86 &    0.575 & B+ & FW  & G  & \\[.9mm]
\ion{Al}{iii} & 4149.91 & 20.55 &    0.626 & A+ & FFTI& C  & 5.99\\
\ion{Al}{iii} & 4149.97 & 20.55 & $-$0.674 & A+ & FFTI& C  & \\
\ion{Al}{iii} & 4150.17 & 20.56 &    0.471 & A+ & FFTI& C  & \\
\ion{Al}{iii} & 4479.89 & 20.78 &    0.900 & X  & KB  & C  & 5.93\\
\ion{Al}{iii} & 4479.97 & 20.78 &    1.020 & X  & KB  & C  & \\
\ion{Al}{iii} & 4480.01 & 20.78 & $-$0.530 & X  & KB  & C  & \\
\ion{Al}{iii} & 4512.57 & 17.81 &    0.408 & A+ & FFTI& C  & 5.96\\
\ion{Al}{iii} & 4528.95 & 17.82 & $-$0.291 & A+ & FFTI& C  & 5.83\\
\ion{Al}{iii} & 4529.19 & 17.82 &    0.663 & A+ & FFTI& C  & \\[.9mm]
\ion{Si}{ii}  & 3862.60 &  6.86 & $-$0.757 & C+ & NIST& C  & 7.08\\
\ion{Si}{ii}  & 4128.05 &  9.84 &    0.359 & B  & NIST& LDA& 7.15\\[.9mm]
\ion{Si}{iii} & 4552.62 & 19.02 &    0.292 & B+ & FFTI& C  & 7.16\\
\ion{Si}{iii} & 4567.84 & 19.02 &    0.068 & B+ & FFTI& C  & 7.19\\
\ion{Si}{iii} & 4574.76 & 19.02 & $-$0.409 & B  & FFTI& C  & 7.25\\
\ion{Si}{iii} & 4716.65 & 25.33 &    0.491 & B  & NIST& C  & 7.15\\[.9mm]
\ion{Si}{iv}  & 4116.10 & 24.05 & $-$0.110 & A+ & FFTI& D91& 7.17\\[.9mm]
\ion{S}{ii}   & 4162.67 & 15.94 &    0.78  & D  & NIST& C  & 7.19\\[.9mm]
\ion{S}{iii}  & 3985.92 & 18.29 & $-$0.79  & E  & WSM & C  & 7.22\\
\ion{S}{iii}  & 4361.47 & 18.24 & $-$0.39  &D$-$& WSM & C  & 7.14\\
\ion{S}{iii}  & 4364.66 & 18.32 & $-$0.71  & E  & WSM & C  & 7.14\\[.9mm]
\ion{Ar}{ii} & 4426.001 & 16.75 &    0.195 & B+ & FFTI& C  & 6.65\\ 
\ion{Ar}{ii} & 4735.905 & 16.64 & $-$0.096 & B+ & FFTI& C  & 6.71\\[.9mm]
\ion{Fe}{iii} & 4081.01 & 20.63 &    0.372 & X  & KB  & C  & 7.35\\
\ion{Fe}{iii} & 4164.73 & 20.63 &    0.923 & X  & KB  & C  & 7.31\\
\ion{Fe}{iii} & 4310.36 & 22.87 &    1.156 & X  & KB  & C  & 7.29\\
\ion{Fe}{iii} & 4310.36 & 22.87 &    0.189 & X  & KB  & C  & \\
\ion{Fe}{iii} & 4372.04 & 22.91 &    0.585 & X  & KB  & C  & 7.33\\
\ion{Fe}{iii} & 4372.10 & 22.91 &    0.029 & X  & KB  & C  & \\
\ion{Fe}{iii} & 4372.13 & 22.91 &    0.727 & X  & KB  & C  & \\
\ion{Fe}{iii} & 4372.31 & 22.91 &    0.865 & X  & KB  & C  & \\
\ion{Fe}{iii} & 4372.31 & 22.91 &    0.193 & X  & KB  & C  & \\
\ion{Fe}{iii} & 4372.50 & 22.91 &    0.200 & X  & KB  & C  & \\
\ion{Fe}{iii} & 4372.54 & 22.91 &    0.993 & X  & KB  & C  & \\
\ion{Fe}{iii} & 4372.78 & 22.91 &    0.040 & X  & KB  & C  & \\
\ion{Fe}{iii} & 4372.82 & 22.91 &    1.112 & X  & KB  & C  & \\
\ion{Fe}{iii} & 4419.60 &  8.24 & $-$2.218 & X  & KB  & C  & 7.25\\
\ion{Fe}{iii} & 4431.02 &  8.25 & $-$2.572 & X  & KB  & C  & 7.30\\[.2mm]
\hline
\end{longtable}
\vspace{-2.5mm}
\tablefoot{
$\varepsilon(X)$\,=\,$\log$\,X/H\,$+$\,12.~~~
Accuracy indicators -- uncertainties within: 
A: 3\%; 
B: 10\%; 
C: 25\%; 
D: 50\%; 
E: larger than 50\%; 
X: unknown.\\[.2mm]
Sources of $gf$-values -- 
BB89:  \citet{BeBu89};
FFT:   \citet{FFT04}; 
FFTI:  \citet{FFTI06};
FW:    \citet{FuWi98}; 
KB:    \citet{KuBe95};
MAR:   \citet{Maretal00};
NIST:  \citet{Kramidaetal14}; 
N02:   \citet{Nahar02};
WFD:   \citet{WFD96};
WSM:   \citet{WSM69}.\\[.2mm]
Broadening data --
C:   approximation formula by \citet{Cowley71};
D91: \citet{Dimitrijevicetal91};
G:   \citet{Griem64,Griem74};
LDA: \citet{Lanzetal88}.
}
\end{longtab}

\renewcommand{\thefigure}{\arabic{figure}\alph{subfig}}
\setcounter{subfig}{1}
\begin{figure*}
\centering
\includegraphics[width=.92\textwidth]{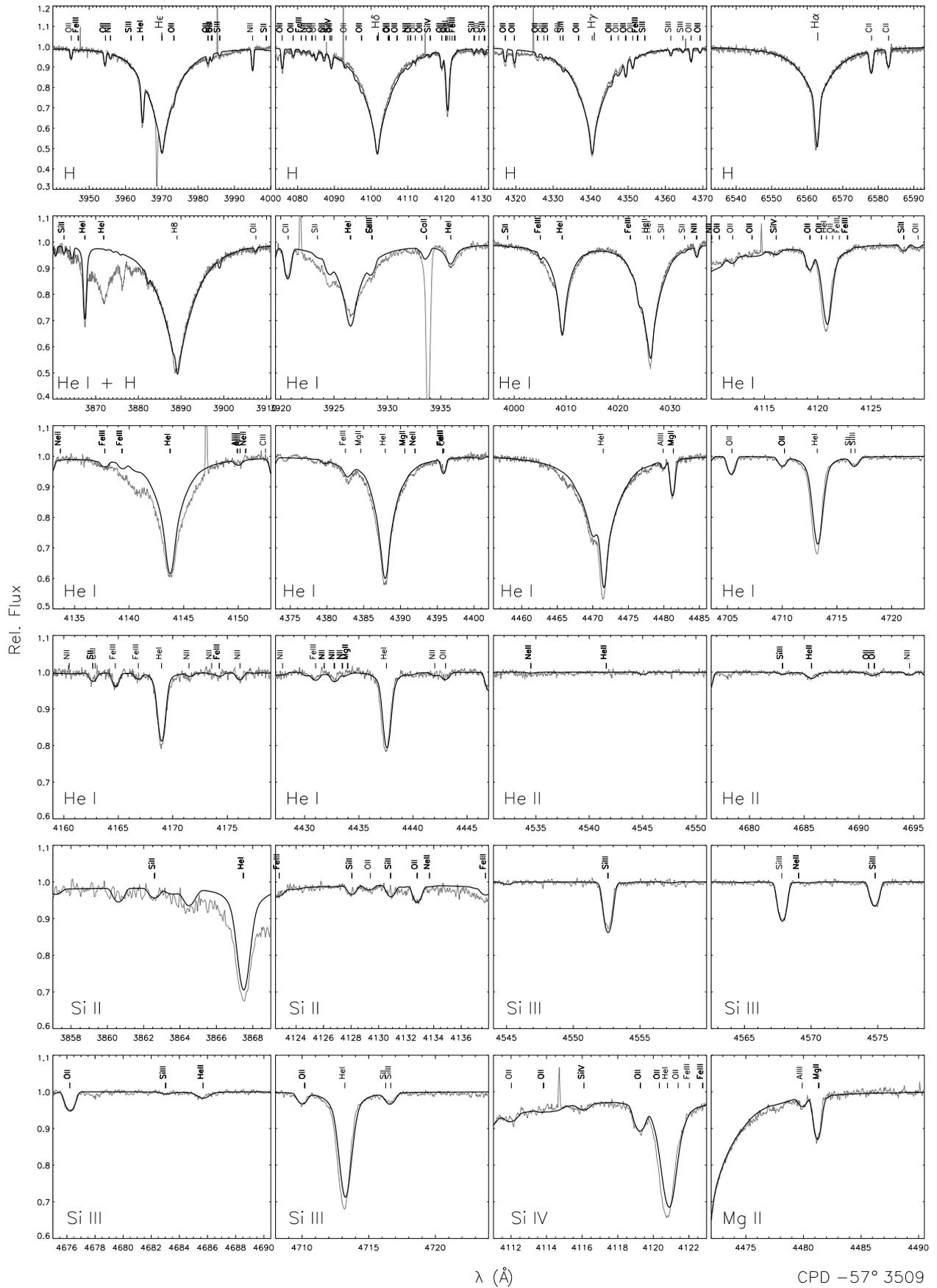}\\[-5mm]
\caption{Comparison of our global best-fit model (thick 
black line) to the normalised FLAMES/GIRAFFE spectrum of CPD\,$-57^{\circ}$\,3509 (grey
line), concentrating here on H, \ion{He}{i/ii}, \ion{Si}{ii/iii/iv} and 
\ion{Mg}{ii} lines. 
The panels are centred on analysed lines, sorted according to chemical
species (as indicated in the lower left of each panel), and, 
within a species, sorted along increasing wavelength. Line identifications
are given.}\label{fig:fitsHHe}
\end{figure*}

\onlfig{
\addtocounter{figure}{-1}
\addtocounter{subfig}{2}
\begin{figure*}
\centering
\includegraphics[width=.92\textwidth]{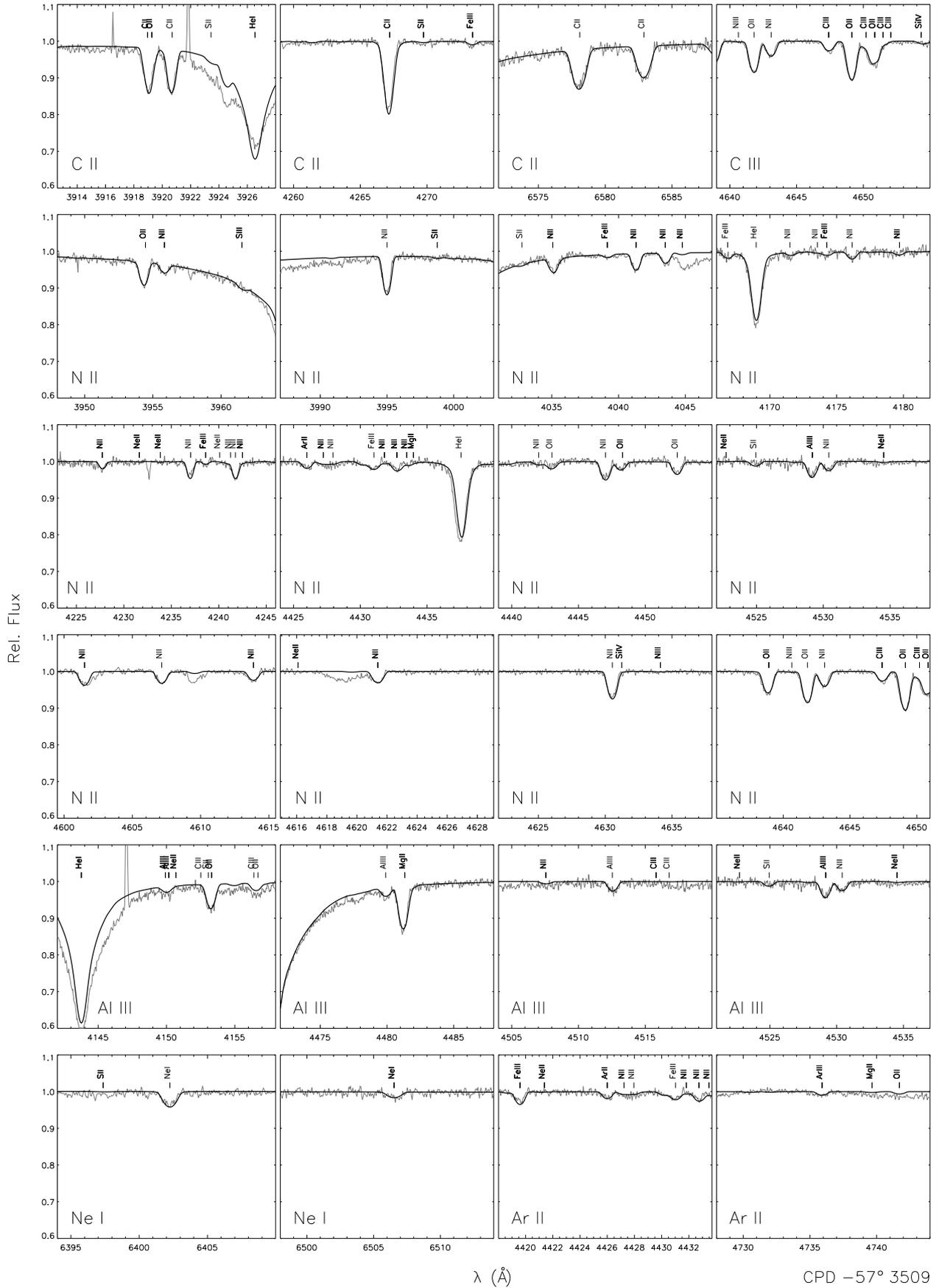}\\[-5mm]
\caption{Same as Fig.~\ref{fig:fitsHHe}, but for spectral lines of
\ion{C}{ii/iii}, \ion{N}{ii}, \ion{Al}{iii}, \ion{Ne}{i} and
\ion{Ar}{ii}.}\label{fig:fitsMET}
\end{figure*}
\clearpage

\addtocounter{figure}{-1}
\addtocounter{subfig}{1}
\begin{figure*}
\centering
\includegraphics[width=.92\textwidth]{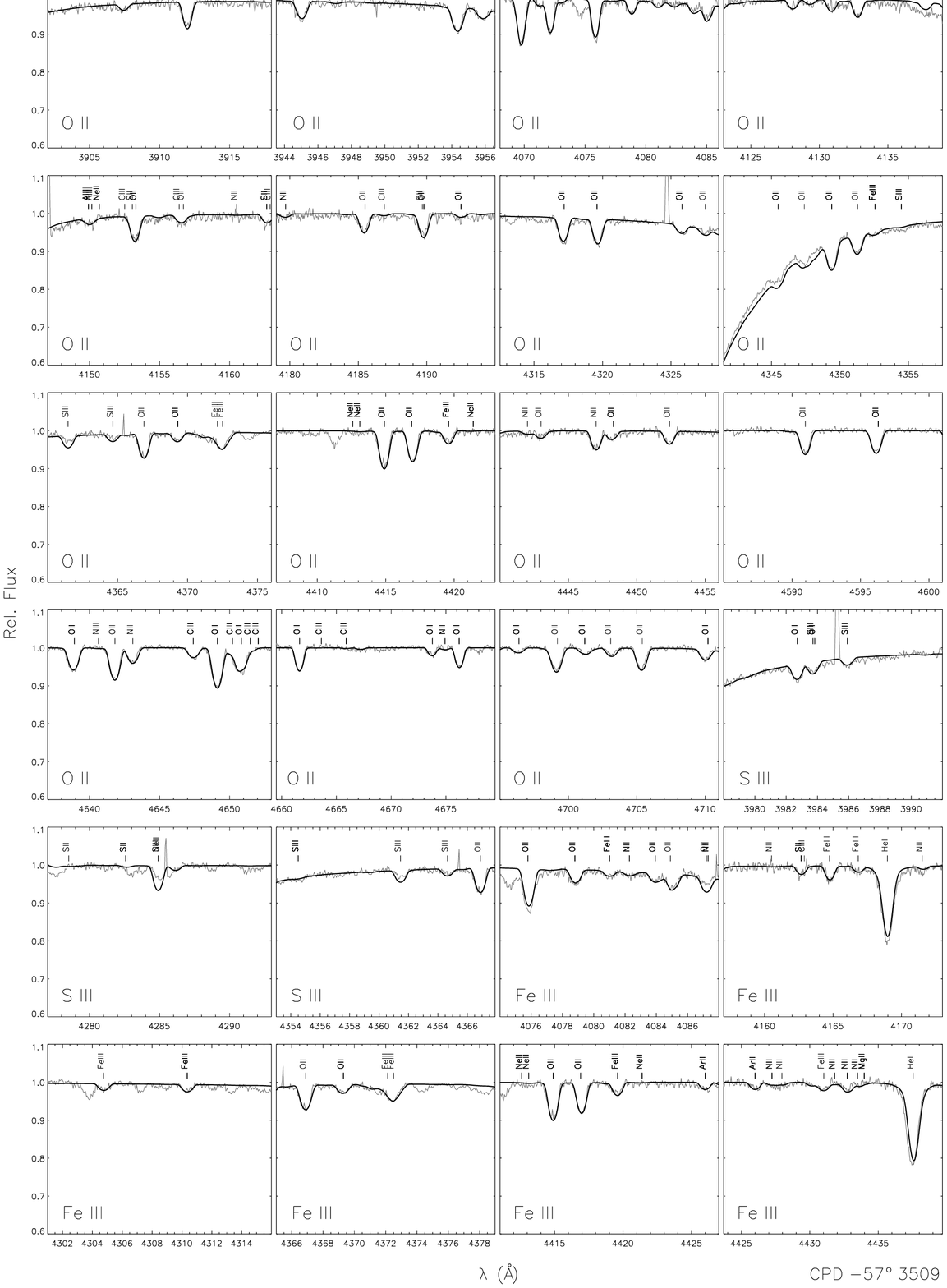}\\[-5mm]
\caption{Same as Fig.~\ref{fig:fitsHHe}, but for spectral lines of
\ion{O}{ii}, \ion{S}{ii/iii} and \ion{Fe}{iii}.}\label{fig:fitsMET2}
\end{figure*}
\clearpage
}

\begin{figure}[t]
\addtocounter{figure}{-1}
\setcounter{subfig}{0}
\centering
\includegraphics[width=87mm,clip]{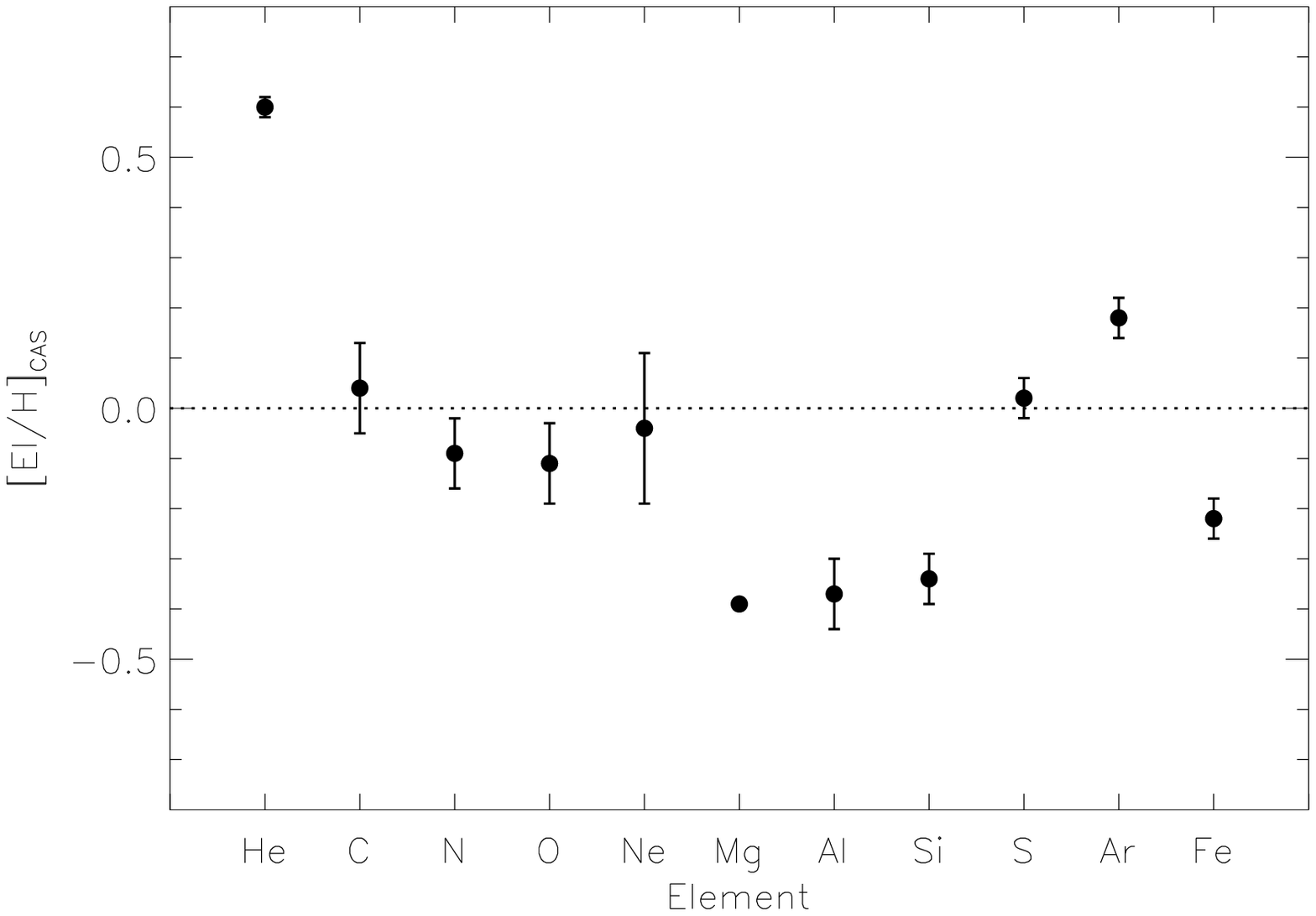}
\caption{Elemental abundances in the atmosphere of CPD\,$-57^{\circ}$\,3509
relative to the cosmic abundance standard (CAS). The error bars show
statistical 1$\sigma$-uncertainties from the line-to-line scatter in each element.
Systematic uncertainties of the abundances amount to about 0.1\,dex.}
\label{fig:cpdabundancesvscas}
\end{figure}

\subsection{Results\label{sect:results}}
Table~\ref{tab:parameters} summarises the results from the
comprehensive characterisation of CPD\,$-57^{\circ}$\,3509 as obtained  in
this work. A first block of entries gives the spectral type, the
barycentric radial velocity $v_\mathrm{rad}$ \citep[which is slightly variable
among the available high-resolution spectra because of the presence of
spots, and broadly consistent with the cluster $v_\mathrm{rad}$ of
12$\pm$5\,km\,s$^{-1}$ as derived by][]{Evansetal05}, the spectroscopic
distance $d_\mathrm{spec}$ (following NP12), and the dipolar field
strength $B_\mathrm{d}$ (assuming a dipolar field configuration,
see Sect.~\ref{sect:discussion} for a discussion). Then, the second and third blocks summarise
the results of the quantitative analysis on atmospheric parameters
and non-LTE metal abundances\footnote{We keep the usual abundance
scale relative to hydrogen here (despite the chemically unusual
overall composition), since this facilitates a comparison to the pristine
composition. The fractionated stellar wind (see Sect.~\ref{sect:discussion} 
for a discussion) couples hydrogen and the metals, while the backfalling helium 
dilutes the metal abundances, apparently reducing the overall metallicity.} 
by number (the number of analysed lines is also indicated), respectively.
Detailed information on the abundances derived for all diagnostic metal 
lines is given in Table~\ref{tab:linelist}, which is only available online. 
There, information on the transition wavelength $\lambda$, the excitation
potential $\chi$ of the lower level of the transition, the oscillator
strength $gf$, an indicator of its accuracy, the source of the
$gf$-value, and a reference for the quadratic Stark broadening data
employed for the computations is also summarised. The final elemental
abundances were calculated giving equal weight to all lines from all
ions of a chemical species, the uncertainties in
Table~\ref{tab:parameters} representing the 1$\sigma$ standard
deviations from the line-to-line scatter. Systematic errors
due to factors like uncertainties in stellar parameters, continuum setting, and 
atomic data on the elemental abundances are difficult to 
quantify accurately \citep[see e.g.][]{Sigut99,Przybillaetal01a,Przybillaetal01b}.
From these previous experiences, we expect them to amount to about 0.1\,dex.

A comparison of the synthetic spectrum based on the parameters from
Table~\ref{tab:parameters} with the observed FLAMES/GIRAFFE spectrum is shown in
Figs.~\ref{fig:fitsHHe}-\ref{fig:fitsMET2} (the last two available
in the online version of the paper) on a line-by line basis.
Figure~\ref{fig:fitsHHe} concentrates on the hydrogen Balmer lines,
the unusually strong helium lines that are characteristic of the
He-strong stars, and the \ion{Si}{ii-iv} lines. Overall, a good fit
between model and observation is achieved. The unmodelled narrow extra absorption
close to the core of H$\epsilon$ is the interstellar Ca H line. Some
slightly blueshifted excess absorption is noticeable mostly in 
several of the \ion{He}{i} lines. This feature is likely due to
a spot on the surface. (Similar signatures can also be seen on a much smaller
scale in some metal lines.) It is not
clear whether vertical chemical stratification is an issue here, because clear-cut
signatures such as core-wing anomalies \citep[e.g.][]{Mazaetal14} are
absent. We note that the poor fit to some of the \ion{He}{i} lines like
$\lambda\lambda$\,3926\,{\AA} and 4141/4143\,{\AA}, or even their absence
from the model ($\lambda$\,3871\,{\AA}), is because of the lack of 
appropriate broadening data in the literature.

A proper determination of the atmospheric indicators is further indicated by
the match of four ionization balances simultaneously, \ion{He}{i/ii},
\ion{C}{ii/iii}, \ion{Si}{ii-iv,} and \ion{S}{ii/iii}
(Figs.~\ref{fig:fitsHHe}-\ref{fig:fitsMET2}). The strongest
constraints stem from silicon, which covers both the main and the
two adjacent minor ionisation stages. The presence of
\ion{He}{ii} at such a low $T_ \mathrm{eff}$ in a main-sequence star
is unusual, but results from the high helium abundance.
The overall good to excellent match of the metal line profiles by
theory -- also for species where only lines from one ion are observed --
reflects the small dispersion in abundances. To our knowledge, this
represents the most comprehensive non-LTE abundance study of any
He-strong star to date.

\begin{figure}[t]
\centering
\includegraphics[width=88mm,clip]{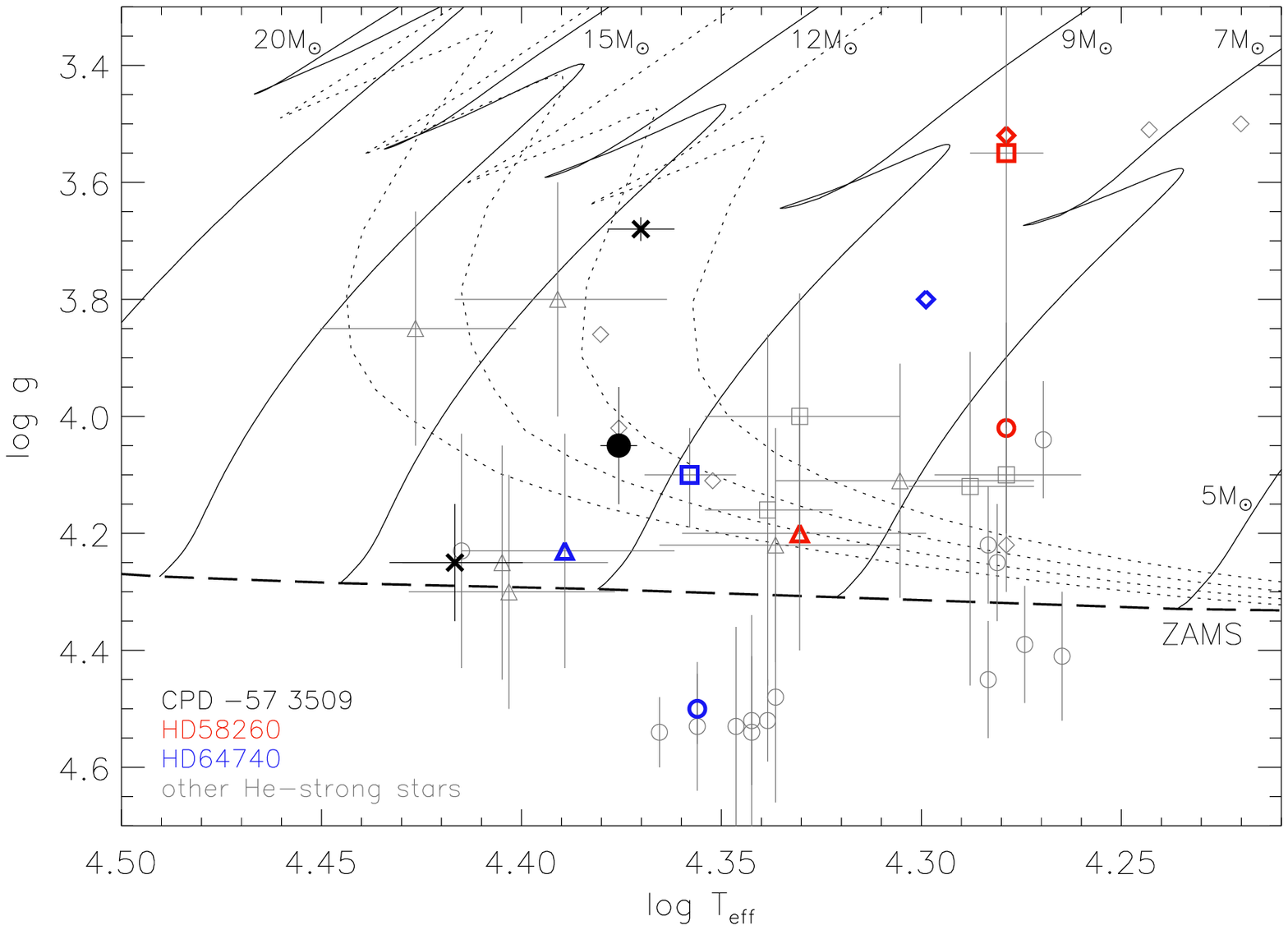}
\caption{CPD\,$-57^{\circ}$\,3509 in the $\log T_\mathrm{eff}$-$\log g$
diagram (black dot, this work). Previous analysis results of the star
obtained by use of grids assuming solar helium abundance
are indicated by St.\,Andrew's crosses, from the work by 
\citet[high-gravity solution]{Hunteretal09} and 
\citet[low-gravity solution]{McSwainetal09}.
Analysis results for other He-strong stars from
the literature are also shown from \citet[circles]{Zboriletal97},
\citet[diamonds]{Leoneetal97}, \citet[squares]{HuGr99}, and
\citet[triangles]{Cidaleetal07}, marked in grey. Two stars common to
all four analyses are highlighted, \object{HD\,58260} (red symbols) and
\object{HD\,64740} (blue symbols). Where available, 1$\sigma$ error bars are displayed.
Overlaid are Geneva evolution tracks for rotating stars, computed for metallicity
$Z$\,=\,0.014 and an initial value of 40\% of the critical rotation
rate \citep[full lines]{Ekstroemetal12}. Corresponding isochrones for $\log
\tau$\,(Myr)\,=\,7.0, 7.1, 7.2, and 7.3 are also displayed (dotted
lines), as well as the position of the zero-age main sequence (ZAMS,
long-dashed line).}
\label{fig:cpdtefflogg}
\end{figure}

The resulting abundance pattern with respect to the cosmic abundance
standard (CAS), as established from early B-type stars in the solar
neighbourhood \citep[preliminary values for Al, S and Ar are adopted
from the latter work]{NiPr12,Przybillaetal13}, is presented in
Fig.~\ref{fig:cpdabundancesvscas}. The common bracket notation is
used: $[$El/H$]$\,=\,$\log$\,(El/H)\,$-$\,$\log$\,(El/H)$_\mathrm{CAS}$.
Besides the high value for helium, a bimodal behaviour is found for the metal
abundances. A group of elements (C, N, O, Ne, S, Ar) shows abundances
close to CAS values, while another group (Mg, Al, Si, Fe) is deficient
by a factor $\sim$2. The first group of elements
seems to be close to the pristine abundances of the NGC\,3293 cluster,
which lies at a Galactocentric distance only $\sim$300\,pc smaller
than that of the Sun; i.e., CAS values should be representative in the absence
of azimuthal abundance variations.

The fourth block of Table~\ref{tab:parameters} concentrates on
photometric data for CPD\,$-57^{\circ}$\,3509. The observed Johnson $V$ magnitude 
and $B-V$ colour are adopted from \citet{Delgadoetal11}. The colour
excess $E(B-V)$ was determined by comparison with synthetic photometry
from the {\sc Atlas9} flux. Correction for extinction (assuming a
ratio of total-to-selective extinction $R_V$\,=\,3.1) allows the
absolute visual magnitude $M_V$ to be derived for the spectroscopic
distance, and application of the bolometric correction from the 
{\sc Atlas9} model allows the bolometric magnitude $M_\mathrm{bol}$
to be determined.

Finally, the fundamental stellar 
parameters mass $M$, radius $R$, luminosity $L$,
evolutionary age $\tau$ and fractional main-sequence lifetime
$\tau/\tau_\mathrm{MS}$ were derived via comparison with stellar
evolution models from the Geneva group \citep{Ekstroemetal12},
as summarised in the last block of Table~\ref{tab:parameters}. 
The location of the star in the $\log T_\mathrm{eff}$-$\log g$
plane (Fig.~\ref{fig:cpdtefflogg}) with respect to the evolutionary tracks and isochrones provided 
the evolutionary mass, age, and fractional main-sequence age, respectively. 
Luminosity and radius followed, once the (spectroscopic) distance was determined.
A second independent derivation of the fundamental parameters 
employed the Bayesian statistical tool {\sc
Bonnsai}\footnote{The {\sc Bonnsai} web-service is available at
{\tt http://www.astro.uni-bonn.de/stars/bonnsai}.} 
\citep{Schneideretal14} on the basis of stellar evolution tracks 
by \citet{Brottetal11}. A Salpeter mass function was adopted as 
prior for the initial stellar masses, a Gaussian rotational velocity 
distribution with mean of 100\,km\,s$^{-1}$ and FWHM of 250 km\,s$^{-1}$
\citep[cf.][]{Hunteretal08} and a uniform prior in age. Furthermore, 
it was assumed that all rotation axes are randomly oriented in space. 
The offset between the two solutions -- though
being within the mutual uncertainties -- is mostly related to the higher
overshooting value adopted in the \citet{Brottetal11} models.
For test purposes, a bi-modal rotational velocity distribution as
derived by \citet{Duftonetal13} was also employed as prior in the 
{\sc Bonnsai} modelling, resulting in practically identical
fundamental stellar parameters as reported in Table~\ref{tab:parameters}. 
For further tests with  {\sc Bonnsai}, we also assumed the star to be an intrinsic faster
rotator, adopting $v_\mathrm{rot}$\,=\,200$\pm$50\,km\,s$^{-1}$.
A slight trend is found that we would overestimate the age and underestimate 
the mass of the star in that case, however with the changes covered
well by the uncertainties stated in Table~\ref{tab:parameters}.

The fundamental stellar parameters may be subject 
to some additional systematic error.
This is because of the chemically normal interior of the star (see
Sect.~\ref{sect:discussion}) and the chemically peculiar atmosphere,
which is not reflected by the evolution models; i.e., the surface
parameters predicted by the models are slightly different than those
observed. However, we expect this to be a secondary effect only, which
is confirmed by the rather good match of the stellar values 
with the NGC\,3293 cluster distance and age of $\sim$2460\,pc and
10.7\,Myr as discussed, for example, by \citet{LoMa94}. 

\begin{figure}[t]
\centering
\includegraphics[width=89mm,clip]{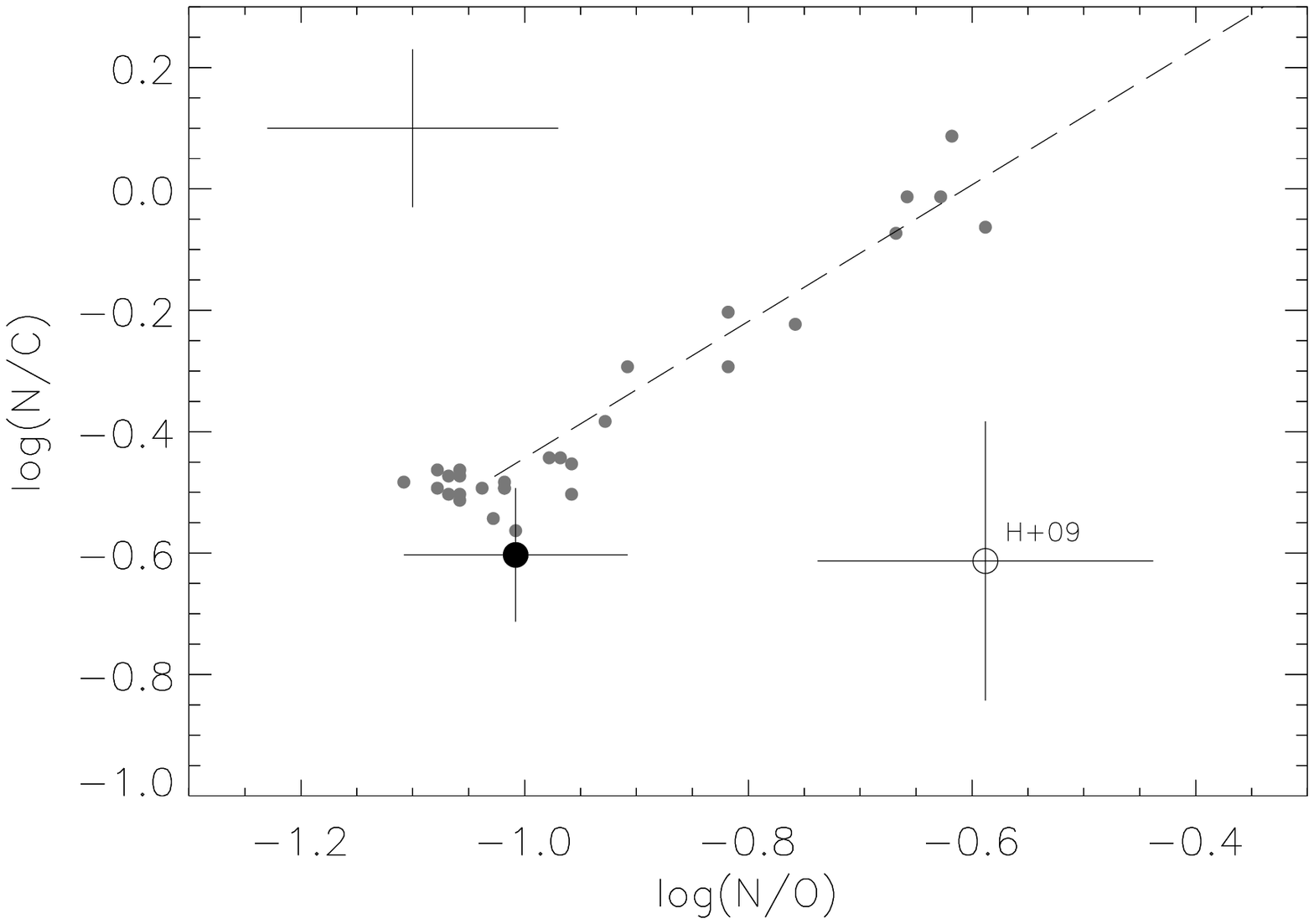}
\caption{Position of CPD\,$-57^{\circ}$\,3509 in the N/O-N/C diagram (by
mass): present work (black dot) and the result from \citet[open
circle]{Hunteretal09}, compared to 29 early B-type stars from
\citet{NiSi11} and NP12 (grey dots). An error bar typical for the latter is indicated in the
upper left corner. The dashed line indicates the analytical approximation to
the nuclear path for CN-cycled material using CAS values \citep{Przybillaetal10}.}
\label{fig:cnopath}
\end{figure}

\subsection{Comparison with previous work\label{sect:previouswork}}
The same FLAMES/GIRAFFE spectrum of CPD\,$-57^{\circ}$\,3509 was analysed previously by
\citet{Hunteretal09}. Their $T_\mathrm{eff}$ of 26\,100$\pm$1000\,K
and $\log g$\,=\,4.25$\pm$0.10 differ significantly from our solution,
a result of the assumption of a solar helium
abundance in their analysis. All other atmospheric parameters and the
metal abundances in common (C, N, Mg, Si) are rather close to our results, 
which is very likely an effect of cancellation of several factors. An 
exception is oxygen, for which they indicate an abundance lower by more than
0.5\,dex. 

Confidence in our solution also comes from a better match 
of the position of CPD\,$-57^{\circ}$\,3509 in the N/C vs. N/O diagram with
respect to the predicted nuclear path, which provides a powerful
quality test for observational results
\citep{Przybillaetal10,Maederetal14}, see Fig.~\ref{fig:cnopath}.
The current position of CPD\,$-57^{\circ}$\,3509 in the diagram may not
reflect its pristine position exactly because of the fractionated stellar wind.
However, we expect systematic effects on the CNO ions to be small
because of their similar atomic structures, similar atomic weights, and 
double-ionization energies near or above the \ion{He}{i} edge; 
i.e.,~their relative abundances should not change much.

CPD\,$-57^{\circ}$\,3509 was also analysed by \citet{McSwainetal09}, based on
lower-resolution spectra ($R$\,$\approx$\,1500--4000). They found
$T_\mathrm{eff}$\,=\,23\,450$\pm$450\,K and $\log g$\,=\,3.68$\pm$0.02
from interpolation of the {\sc Tlusty} BSTAR2006 grid of \citet{LaHu07} 
aiming at minimisation of the root-mean-square difference between model 
and observation for a single line indicator, H$\gamma$. Their
$v\sin i$\,=\,62$\pm$13\,km\,s$^{-1}$ is a result of concentrating on
\ion{He}{i} lines in their analysis, which resulted in a higher $v\sin
i$-value because the He-strong nature of the star was not recognised.
The positions of CPD\,$-57^{\circ}$\,3509 for both the solutions by 
\citet{McSwainetal09} and \citet{Hunteretal09} are indicated in
Fig.~\ref{fig:cpdtefflogg}.

\section{Discussion\label{sect:discussion}}
Between the FORS2 (considering the whole spectrum) and HARPSpol 
observations, the highest \bz\ value in modulus was obtained from the 
analysis of the FORS2 data obtained on 2 June 2014. 
Considering then \bz$_\mathrm{max}$\,$\sim$\,1\,kG and assuming a dipolar configuration 
of the magnetic field, we derive a lower limit on the dipolar field 
strength of B$_\mathrm{d}$\,$\gtrsim$\,3.3\,kG, following
\citet[their Sect.~2.7]{Auriereetal07}. 
In addition, the large and fast (within about 1 day) variation in 
the \bz\ value shown by the analysis of the FORS2 data indicates that the star is 
rotating faster than implied by the low $v \sin i$-value 
with a likely rotation period of a few days or less.
An estimate can be achieved from $v_\mathrm{rot}$\,=\,$2\pi R/P$,
assuming a rotation period $P$ in the range one to three\,days and adopting
the range in radius, 4.4 to 5.0\,$R_\odot$, obtained from considering 
the Geneva and Bonn models. One finds $v_\mathrm{rot}$\,$\approx$\,70
to 250\,km\,s$^{-1}$, so less than about 50\% of the critical velocity.
This may be viewed as an upper limit since the magnetic field
geometry could be more complex than dipolar. An inclination angle 
$i$ in the range $\sim$8 to 30$^{\circ}$ is implied.

From the magnetic field variations, we get values between $-$560\,G and
+1050\,G (from the Bonn reduction). Since we do not have the fully
modulated magnetic field variation curve, we do not know the 
real \bz$_\mathrm{min}$ and \bz$_\mathrm{max}$. But if we use the
measured values, we get \bz$_\mathrm{min}$/\bz$_\mathrm{max}$\,$=$\,$-$0.53,
 $\beta$\,=\,88$^{\circ}$ for $i$\,=\,8$^{\circ}$, and
$\beta$\,=\,80$^{\circ}$ for $i$\,=\,30$^{\circ}$, with
$\beta$ being the obliquity angle, i.e. the angle between the
rotation axis and the magnetic axis (see e.g. Sect.~3 of
\citet{Hubrigetal11b} for the formalism employed here).
Of course, if the maximum/minimum values of \bz\ differ significantly from the 
assumed values, $\beta$ will also change dramatically. And if we were looking
equator-on, meaning that $\sin i$\,=\,1, it would be more difficult to make a
statement about $\beta$. However, a $\beta$ close to 0$^{\circ}$ is highly unlikely.
To summarise here, a low inclination value combined with a high obliquity angle offers a
plausible explanation for the apparent contrast between the small
width of the absorption lines and the short timescale of the variation
in the observed magnetic field strength.

In the context of the classification of magnetospheres of massive
stars presented by \citet{petit2013}
and assuming a minimum dipolar magnetic field strength of 3.3\,kG,
we obtained a lower limit on the Alfv{\'e}n radius of about 35 stellar radii and
an upper limit on the Keplerian corotation radius of about 7 stellar
radii\footnote{\citet{petit2013} do not
cover the case of oblique rotators, which is highly relevant here, adding
a further degree of uncertainty to the discussion.}. 
For the calculation of the Alfv{\'e}n and Keplerian corotation
radius we adopted the stellar parameters obtained from {\sc Bonnsai}
as presented in Table~\ref{tab:parameters}, a terminal velocity of 700\,$\mathrm{km\,s}^{-1}$, and
mass-loss rate $\dot{M}$  in the range of 10$^{-11}$ to 10$^{-10}$\,$M_\odot$\,yr$^{-1}$ 
typical of weak winds of magnetic B dwarfs \citep{Oskinovaetal11}.

The derived values indicate that the star should be able to support a centrifugal
magnetosphere. The star could therefore hold a circumstellar disk or
cloud formed of stellar wind material trapped within the magnetic
field lines, which usually reveals itself by emission, for example in
the H$\alpha$ line. The spectra collected so far do not
present any signature indicative of the presence of circumstellar
material. CPD\,$-57^{\circ}$\,3509 therefore seems to be one of the
examples of He-strong stars with H$\alpha$ absorption, similar to
HD\,58260 \citep{Pedersen79} or \object{HD\,96446} \citep{Neineretal12}, 
which indicate that a centrifugal magnetosphere is a necessary 
but not sufficient condition for developing emission.

 Apparently, the
wind is either not strong enough for sufficient material to accumulate in the
magnetosphere to become observable, or alternatively, some leakage
process leads to loss of material from the magnetosphere (see
\citet{petit2013} for further discussion). One may speculate that the
magnetic field topology, in particular deviations from a global dipole
that can favour magnetic reconnection, may
play a r\^ole in such a leakage process by weakening the magnetic
confinement of the circumstellar material. Alternatively, a high
obliqueness of the magnetic field with respect to the rotation axis may 
inhibit the formation of a circumstellar disk, since mass loss into some
solid angle can occur along field lines in the equatorial plane in
that case.

Once considered a small number of oddballs, it is now
clear that He-strong stars compose an important second class of 
magnetic objects among massive stars, easily identified 
spectroscopically like the magnetic Of?p stars.
Their quantitative analysis, like that for any chemically-peculiar 
object, is more demanding than for ordinary stars. That said, 
it seems that the atmospheric parameters of the He-strong stars 
are rather poorly constrained, since huge systematics between different 
studies of the same star are apparent (see Fig.~\ref{fig:cpdtefflogg}).
Essentially, any position from close to the zero-age main sequence
(ZAMS) to near the end of core hydrogen burning, and even below the
ZAMS, has been assigned to some prototype objects of this class in several
previous studies.
Also, differences in $T_\mathrm{eff}$ can be considerable.
Further studies with modern non-LTE modelling techniques, as applied here,
are certainly needed to improve our understanding of this class of star.
In particular, it is imperative to account for the peculiar helium
abundances in the modelling, because its neglect can lead to large
systematic effects on the analysis, see the discussion in
Sect.~\ref{sect:previouswork}. 

The helium abundance of $y$\,=\,0.28 (which corresponds to a mass
fraction of 0.6) in the atmosphere of CPD\,$-57^{\circ}$\,3509 locates 
the star in the upper quartile of the helium abundance distribution for 
He-strong stars \citep{Zboriletal97}\footnote{\citet{Zboriletal97} give LTE helium abundances. Non-LTE effects tend to
strengthen the \ion{He}{i} lines in the optical by a few to several
10\% in equivalent width (depending on line), so that their abundances 
should be viewed as upper limits.}. Given the star's luminosity 
of $\log L/L_\odot$\,$\approx$\,3.8 one can conclude that the
He enrichment is indeed confined to the atmospheric layers, and the envelope
has normal He composition: if the He mass fraction inside the star were 
as high as 0.6, then its luminosity would have to be much higher: 
$\log L/L_\odot$\,=\,4.2, according to the mass-luminosity-helium mass fraction 
relation \citep{Graefeneretal11}.

Non-LTE abundances for all elements with lines in the optical spectra
in early B-stars were derived here for the first time for a He-strong
star. It is therefore worthwhile discussing the resulting abundance 
pattern for CPD\,$-57^{\circ}$\,3509 (Fig.~\ref{fig:cpdabundancesvscas}) further. 
It is probably a consequence of the fractionated 
stellar wind that also gives rise to the peculiarity in helium abundance
\citep{HuGr99,KrKu01}. The weak wind prevailing at the $T_\mathrm{eff}$ 
of CPD\,$-57^{\circ}$\,3509 is radiatively driven by metal species with pronounced 
line spectra longwards of the Lyman jump, i.e. those that can
efficiently absorb momentum from the radiation field near maximum flux. 
Good examples for these are silicon and iron. All other 
elements we analysed show a relatively sparse line density and typically
weak lines, so they participate in the outflow by being
accelerated indirectly via Coulomb collisions, like hydrogen. As they
are typically ionized singly, they do not decouple from the outflow
and fall back to the surface like {\em neutral} helium (part of the helium
remains ionized and consequently gets dragged along with the stellar
wind). More detailed
investigations are required for magnesium and aluminium, but one may
speculate that their observed underabundances may be the result 
of the stronger Coulomb coupling, because they are predominantly
doubly ionized (both also show a few strong lines in the UV).
This picture allows the prediction that other iron-group species that
have an electron configuration with a partially filled 3$d$ valence
shell similar to iron (and similar ionization energies)  
should also show underabundances by a factor $\sim$2 relative to
cosmic values, which could be verified by UV spectroscopy.

One may assume that the abundances of helium and of the metals vary as a 
function of time in He-strong stars as they evolve off the ZAMS. The helium abundance would be
expected to increase with time (due to fall-back), while the metal abundances 
should decrease (due to the fractionated, metal-rich wind). Quantifying the behaviour
of $y(t)$ and $\log$(X/H)\,$[t]$ is not straightforward and would
require computations like those of \citet{Michaudetal87} to be
undertaken, refined by modern input physics, but this is beyond the scope
of the present paper. While observational evidence for this
time-dependent behaviour is weak at best at present \citep[see the discussion
by][]{Zboriletal97} because of large observational uncertainties,
applying the analysis methods presented here to a sample of
He-strong stars may be worthwhile in order to investigate the question once again.

\begin{figure}[t]
\centering
\includegraphics[width=89mm,clip]{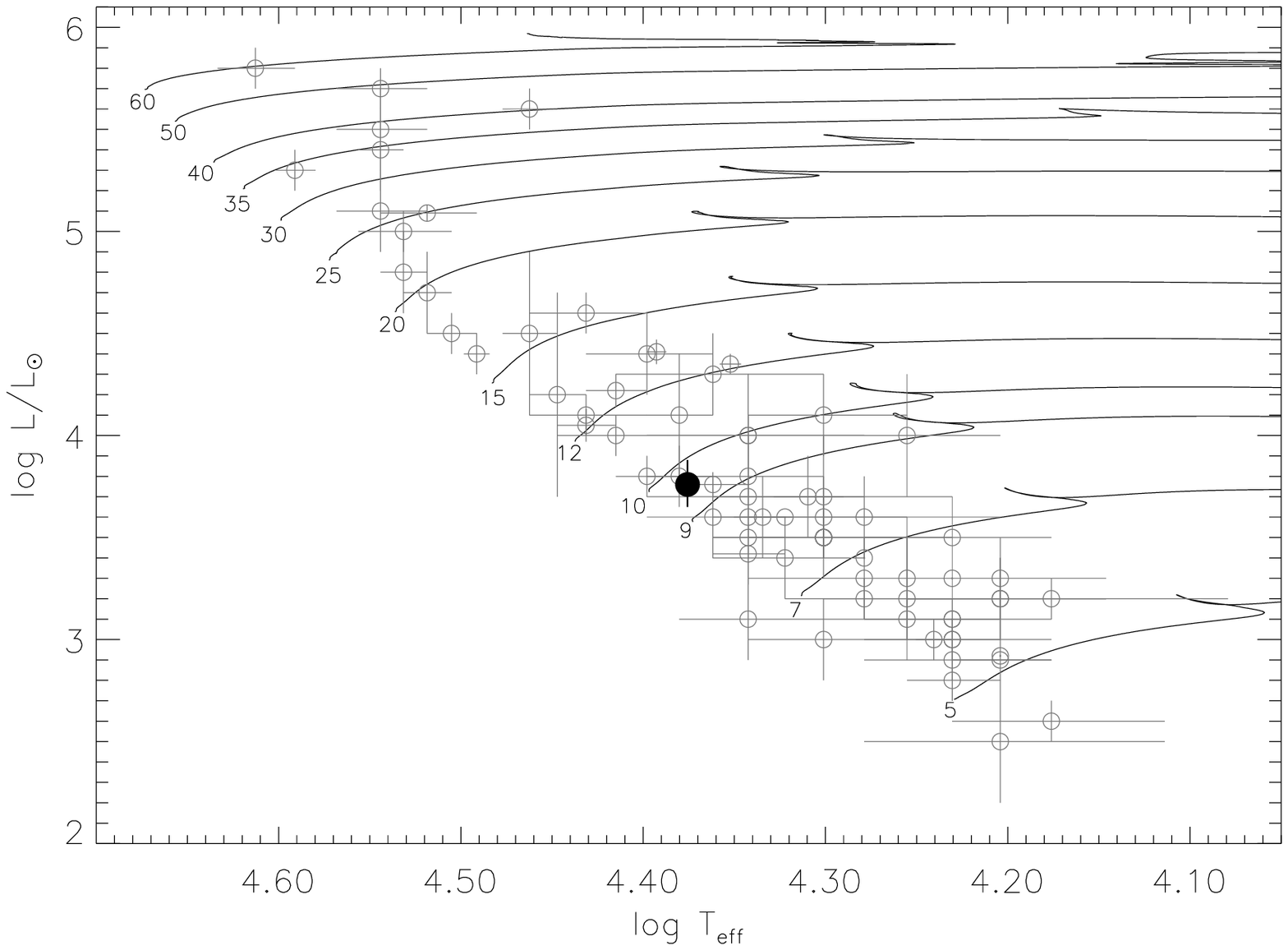}
\caption{Massive stars with confirmed magnetic field detections in the
Hertzsprung-Russell diagram \citep[grey circles]{Briquetetal13,petit2013,Alecianetal14,
Neineretal14,Fossatietal14,Fossatietal15a,Sikoraetal15,Castroetal15},
and CPD\,$-57^{\circ}$\,3509 (black dot). Error bars are 1$\sigma$ uncertainties.
Evolutionary tracks of \citet{Brottetal11} for initial rotational
velocity of $\sim$300\,km\,s$^{-1}$ are shown (full lines), labelled by
initial stellar mass. The magnetic star in the Trifid nebula and
\object{HD\,345439} \citep{Hubrigetal14,Hubrigetal15b} are not 
included here because of difficulties in establishing their properties.}
\label{fig:tefflogl}
\end{figure}

Finally, we want to discuss the evolutionary status of CPD\,$-57^{\circ}$\,3509.
Its position in the Hertzsprung-Russell diagram (HRD) is shown in
Fig.~\ref{fig:tefflogl} with respect to other currently known magnetic massive 
stars. Noteworthy is the good consistency of the star's position 
in the HRD and the $T_\mathrm{eff}$-$\log g$ diagram (Fig.~\ref{fig:cpdtefflogg})
relative to the stellar evolution tracks\footnote{\citet{LaKu14} have shown 
that the position of a star in 
the $T_\mathrm{eff}$-$\log g$ and the HRD diagram differ drastically
only for highly helium-enriched objects (i.e.~showing He-enrichment
not only on the surface). The present consistency indicates that standard
stellar models, i.e.~models neglecting the surface He-enrichment, may be used to
derive fundamental stellar parameters for He-strong stars without inducing 
large systematic uncertainties.}, despite the different sources of the tracks.
It has evolved significantly away from the ZAMS, having burned about
half of its core hydrogen. The star is close to the point where
evolution speeds up towards the terminal-age main sequence and
-- given the uncertainties among previous studies -- among the most 
evolved He-strong stars known, see Fig.~\ref{fig:cpdtefflogg}.
This is consistent with an evolutionary age (see
Table~\ref{tab:parameters}) that compares reasonably well with the 
age of the parent open cluster NGC\,3293, 10.7\,Myr \citep{LoMa94}.
There is also reasonably good agreement between the spectroscopic distance of
CPD\,$-57^{\circ}$\,3509 ($d_\mathrm{spec}$\,=\,2630$\pm$370\,pc) and the 
cluster distance of $\sim$2460\,pc \citep{LoMa94}. By adopting this cluster
distance and the values for reddening and bolometric correction
determined here, one obtains an independent $\log L/L_\odot$\,=\,3.77, 
in excellent agreement with both our values
derived from the \citet{Ekstroemetal12} and \citet{Brottetal11} tracks.

\begin{acknowledgements}
LF acknowledges financial support from the Alexander von Humboldt Foundation.
TM acknowledges financial support from Belspo for contract
PRODEX GAIA-DPAC. FRNS acknowledges the fellowship awarded by the
Bonn–Cologne Graduate School of Physics and Astronomy.
\end{acknowledgements}

%-------------------------------------------------------------------

%\listofobjects

\end{document}